\documentclass[showpacs,preprintnumbers,amsmath,amssymb]{revtex4}
\usepackage{graphicx}
\usepackage{bm}
\usepackage{setspace}
\usepackage{epstopdf}
\usepackage{float}
\usepackage{dcolumn}
\usepackage{bm}
\usepackage{mathrsfs}
\usepackage{amsmath}

\begin{document}

\title{Light bending in a two black hole metric}
\author{M. Alrais Alawadi}
\email{100044354@ku.ac.ae}
\affiliation{%
Department of Mathematics,\\  Khalifa University of Science and Technology,\\ Main Campus, Abu Dhabi,\\ United Arab Emirates}
\author{D. Batic}
\email{davide.batic@ku.ac.ae}
\affiliation{%
Department of Mathematics,\\  Khalifa University of Science and Technology,\\ Main Campus, Abu Dhabi,\\ United Arab Emirates}
\author{M. Nowakowski}
\email{mnowakos@uniandes.edu.co}
\affiliation{
Departamento de Fisica,\\ Universidad de los Andes, Cra.1E
No.18A-10, Bogota, Colombia
}

\date{\today}

\begin{abstract}
We discuss the propagation of light in the C-metric. We discover that null geodesics admit circular orbits only for a certain family of orbital cones. Explicit analytic formulae are derived for the orbital radius and the corresponding opening angle fixing the cone. Furthermore, we prove that these orbits based on a saddle point in the effective potential are Jacobi unstable. This completes the stability analysis done in previous literature and allows us to probe into the light bending in a two black hole metric. More precisely, by constructing a suitable tetrad in the Newmann-Penrose formalism, we show that light propagation in this geometry is shear-free, irrotational, and a light beam passing by a C-black hole undergoes a focussing process. An exact analytic formula for the compression factor $\theta$ is derived and discussed. Furthermore, we study the weak and strong gravitational lensing when both the observer and the light ray belong to the aforementioned family of invariant cones. In particular, we obtain formulae allowing to calculate the deflection angle in the weak and strong gravitational lensing regimes.

\noindent

\end{abstract}

\maketitle
\newpage
\pagenumbering{arabic}

\section{Introduction}
According to Einstein, the presence of a mass deforms the space-time geometry in such a way that light rays passing nearby get deflected. This effect was observed for the first time in 1919 when the English physicist Arthur Eddington, taking advantage of a total eclipse of the Sun, measured the position of some stars that were very close to the edge of the Sun. With surprise, the stars appeared displaced by a small amount compatible with that predicted by General Relativity. The mass of the Sun had caused the deflection of the light rays emitted by the stars, thus changing their apparent position. It took little to understand that what Eddington observed was just the least striking aspect of this gravitational distortion and that the phenomenon could be used to probe the cosmic depths with an accuracy never imagined before. This measurement marked the sunset of Newtonian gravitational physics and opened the door to modern cosmology. Gravitational lensing became and still is an extremely active research field. Over the years, the underlying theory took three different paths:
\begin{enumerate}
\item
{\bf{Strong gravitational lensing}} occurs when the gravitational lens is represented by astrophysical objects  exhibiting an extremely strong gravitational field such as black holes, galaxies or galaxy clusters. In these cases, an observer may not only see one image of a light source behind the gravitational lens but several images. The first example of a strong gravitational lens was discovered in 1979 and it is represented by the so-called {\it{Twin Quasar}} Q0957 + 561 \cite{Q}. Another famous example is the Einstein cross, i.e. a fourfold representation of the same source, in the constellation Pegasus discovered in 1985 \cite{Huchra}. Under certain circumstances, the object behind the gravitational lens appears as a closed line in the form of an Einstein ring \cite{Falco}. The first gravitational lens, consisting of a galaxy cluster (Abell 370) rather than a single galaxy, was recognized as such in 1987 by \cite{Soc}.
\item
{\bf{Weak gravitational lensing}} takes place in very different scenarios, such as when the lens is not very compact, the source is not exactly behind the lens, or the lens is compact but the light rays travel at radial distances from the lens that are much larger than the gravitational radius of the lens itself. In this regime, the deflection of light leads to less dramatic consequences, for example only slight distortions in the form of a distant galaxy \cite{Falco}. 
\item
{\bf{Gravitational microlensing}} occurs whenever the spatial extent of the lens is very small compared to the size of the entire lens system. Typical lenses here are planets, stars or other compact objects. The typical observation in this case are magnification effects that manifest themselves in peaks in the light curves of the sources. The first major application of the microlensing effect was the search for the so-called massive astrophysical compact halo objects (MACHOs), which are considered a possible candidate for dark matter \cite{Pac,Hans}. The microlensing effect is also a promising approach when looking for exoplanets \cite{Ken}.
\end{enumerate}
The gravitational lensing in a Schwarzschild background was extensively studied in \cite{Darwin,Falco,Pet,Bozza,Virba,Bis,Bozza1,Iyer,Liu1} while its counterpart in the presence of a positive cosmological constant, i.e. the Kottler or Schwarzschild-de Sitter metric, was taken under scrutiny in \cite{Islam,Bal,Ser0,Rindler,Ser,Rindler1,Ara,usPRD}. The weak and strong lensing for a Reissner-Nordstr\"{o}m black hole has been tackled by \cite{Sereno,Pang1}. Light deflection in the presence of an astrophysical object whose gravitational field is described by the Kerr metric was analyzed in \cite{Epstein,Richter,Bray,Glin,Scarpetta,Liu2} while a Morse-theoretical analysis of the gravitational lensing in the Kerr-Newman metric was done by \cite{Hasse}. Moreover, a rigorous mathematical framework for calculating the propagation of light rays in the stationary post-Newtonian field of an isolated celestial body (or system of bodies) considered as a gravitational lens having a complex multipole structure was developed in \cite{Kop1}. In addition, \cite{Asada} showed that a rotating lens cannot be  distinguished from a non-rotating one if only expansions at the first order in Newton's gravitational constant are considered. This result was further refined in \cite{Asada1}. Gravitational lensing in metric theories of gravity in post-post Newtonian order with gravitomagnetic field was considered by \cite{Sereno2003}. The study of images produced by black holes with thin accretion disk or near a neutron star was performed by \cite{Luminet}. Regarding the construction of an exact lens equation in the Schwarzschild metric and in the case of spherically symmetric and static space-times, we refer to \cite{Virba,Frit,Per}. Gravitational lensing in the presence of spherically symmetric naked singularities have been studied by \cite{V1,Rus,Kee,usPRD}. Since relativistic images in the strong gravitational regime offer the opportunity to test the theory of gravity, there has been a proliferation of studies on the strong lensing for disparate metrics arising from the Einstein theory of gravity \cite{Einstein}, alternative theories \cite{Altt}, noncommutative geometry \cite{usPRD}, and string theory \cite{String}. We bring the above literature review to an end by mentioning that \cite{Amore} developed a novel method to obtain arbitrarily accurate analytical expressions for the deflection angle of light propagating in a generic spherically symmetric static metric by mapping the integral giving the deflection angle into a rapidly convergent series and \cite{Fortunato,Giannoni,Caponio} applied the Fermat principle to study the lensing of null rays in stationary space-times.

In the present work, we offer a detailed study of the light bending in a two black hole metric. To this purpose, it is essential to take under scrutiny the geodesic motion of test particles in a gravitational field, and in particular, the case of massless particles is here of special interest as it includes the motion of light \cite{light}. The question of new phenomena  arises if we consider a metric which in some limiting case reduces to the Schwarzschild metric (for examples see \cite{we1}). Here, the fate of the circular orbit, already appearing in the Schwarzschild metric, and issues regarding its stability deserve careful attention because they will give us useful insights on how to construct an appropriate impact parameter. A candidate for such a field of investigation is the C-metric  of two black holes which we will briefly introduce below. This manifold , originally constructed by Weyl in 1917 \cite{Weyl}, is a special case of the Pleb$\acute{\mbox{a}}$nski-Demi$\acute{\mbox{a}}$nski family of metrics \cite{PodG}. It represents a pair of causally disconnected black holes, each having mass $M$, and accelerating in opposite directions under the action of forces generated by conical singularities \cite{GKP,KW,B}. Its line element in Boyer-Lindquist coordinates and in geometric units ($G=c=1$) is given by \cite{GKP}
\begin{equation}\label{line-ell}
ds^2=F(r,\vartheta)\left[-f(r)dt^2+\frac{dr^2}{f(r)}+\frac{r^2}{g(\vartheta)}d\vartheta^2+r^2 g(\vartheta)\sin^2{\vartheta}d\varphi^2\right]
\end{equation}
with
\begin{equation}\label{Ffg}
F(r,\vartheta)=(1+\alpha r\cos{\vartheta})^{-2},\quad f(r)=(1-\alpha^2 r^2)\left(1-\frac{2M}{r}\right),\quad g(\vartheta)=1+2\alpha M\cos{\vartheta},
\end{equation}
where $\vartheta\in(0,\pi)$, $\varphi\in(-\kappa\pi,\kappa\pi)$, and $r_H<r<r_h$. Here, $r_H=2M$ represents the Schwarzschild horizon, $\alpha$ is the acceleration parameter, and $r_h=1/\alpha$ denotes the acceleration horizon. In order to preserve the spatial ordering of the horizons, we need to assume that $0<2\alpha M<1$. Furthermore, the computation of the Kretschmann invariant for (\ref{line-ell}) signalizes that the only curvature singularity occurs at $r=0$ while the horizons of the C-black hole are mere coordinate singularities. It is interesting to observe that for $\alpha\to 0$ the line element (\ref{line-ell}) goes over into the line element of a Schwarzschild black hole. This fact means that predictions concerning the bending of light in the C-metric should go over for vanishing $\alpha$ into the corresponding predictions in the Schwarzschild metric. In order to remove the conical singularity at $\vartheta=0$, the parameter $\kappa$ entering in the definition of the range for the angular variable $\varphi$ can be chosen to be \cite{GKP}
\begin{equation}
\kappa=\frac{1}{1+2\alpha M}.
\end{equation}
According to \cite{GKP}, the conical singularity with constant deficit angle along the half-axis $\vartheta=\pi$ has the interpretation of a semi-infinite cosmic string under tension extending from the source at $r=0$ to conformal infinity. This allows to interpret (\ref{line-ell}) as a Schwarzschild-like black hole that is being accelerated along the axis $\vartheta=\pi$ by the action of a force corresponding to the tension in a cosmic string. Note that the range of the rotational coordinate can be restored to its usual value $2\pi$ by means of the rescaling $\varphi=\kappa\phi$ leading to $\phi\in(-\pi,\pi)$. In what follows, we will work with the version of the line element (\ref{line-ell}) obtained after rescaling of the variable $\varphi$, namely 
\begin{equation}\label{line-el01}
ds^2=-B(r,\vartheta)dt^2+A(r,\vartheta)dr^2+C(r,\vartheta)d\vartheta^2+D(r,\vartheta)d\phi^2,
\end{equation}
where
\begin{equation}\label{ABCD}
B(r,\vartheta)=f(r)F(r,\vartheta),\quad A(r,\vartheta)=\frac{F(r,\vartheta)}{f(r)},\quad C(r,\vartheta)=r^2\frac{F(r,\vartheta)}{g(\vartheta)},\quad D(r,\vartheta)=\kappa^2 r^2g(\vartheta)F(r,\vartheta)\sin^2{\vartheta}
\end{equation}
with $F$, $f$, and $g$ defined in (\ref{Ffg}). The last four decades witnessed a growing interest in the C-metric. Several different features have been analysed, for instance: singularities and motion \cite{E1,Hog,Bi}, horizon structure \cite{Far,Zim}, spacetime properties at infinity \cite{Ash,Dray,Sla}. Further studies were undertaken by \cite{Letelier} leading to the discovery that an accelerated black hole has Hawking temperature larger than the Unruh temperature of the accelerated frame. Moreover, separability of test fields equations on this background was extensively studied in \cite{Prest,Kof}. Bound orbits of massless particles are intimately connected to the so-called black hole shadow. The analysis of such orbits also known as fundamental photon orbits in a generic space-time was considered in \cite{Pedro}. Further, analysis of such orbits focusing on light rings in other two black hole metrics, more precisely in the double Schwarzschild, Reissner-Nordstr\"{o}m , and Kerr space-time was performed in \cite{Flavio,Pedro1,Pedro2,Grenz}. Finally, the stability analysis of circular orbits for spinning test particles around an accelerating Kerr black hole was done in \cite{ZH}. Concerning geodesic motion in the C-spacetime, \cite{Pravda} studied in detail the geodesic trajectories of time-like particle and in the case of light-like particles, the authors claimed that photon circular orbits are unstable without providing a solid mathematical proof. The reason could be that they based their conclusions on a local maximum in the effective potential. In contrast to this, we found that the circular orbit is due to a saddle point which requires an additional effort to probe into the associated stability problem. Further work on time-like circular orbits and their null limit has been undertaken in \cite{Lim,Bini2,Bini3} where the corresponding stability problem was not studied.  In \cite{Claudel}, the authors studied the concept of photon surfaces and showed that a special case of photon surface is a photon sphere. Furthermore, subject to an energy condition, they were able to prove that any photon sphere must surround a black hole, a naked singularity or more than a certain amount of matter provided that the underlying manifold is static and spherically symmetric. Moreover, \cite{Bini3} discovered the existence of an invariant plane for the motion of photons in the C-metric but the authors neither showed that a photon circular orbit is allowed nor they derived an analytical formula for the radius of a circular orbit orbit. Moreover, the stability problem of motion on  the invariant plane is nowhere addressed in the literature which seems to us necessary as the orbit emerges from a saddle point in the effective potential. We fill this gap in the next section where we analyse the (in)stability of the photon circular orbits by constructing the second KCC (Kosambi-Cartan-Chern) invariant \cite{21,22,23,28,H} associated to the system of geodesic equations describing the motion of a light-like particle in the C-metric. For a detailed exposition of the Jacobi stability analysis of dynamical systems and applications in gravitation and cosmology we refer to the work of \cite{Baha,Bom,Bom1,Harko1,Danila,Lake}.

According to a theorem in \cite{1,6}, the circular orbits of null particles are Jacobi stable if and only if the real parts of the eigenvalues of the second KCC invariant are strictly negative everywhere (along the trajectory), and Jacobi unstable, otherwise. The result emerging from our analysis is that such trajectories are Jacobi unstable, and therefore, we completed the stability analysis done in previous literature.  The purpose of our work is twofold: we study the gravitational lensing in the presence of a two black hole metric and explore how it differs from the Schwarzschild black hole lensing. In particular, the weak gravitational lensing on a certain family of invariant cones emerging from the stability analysis enables us to describe the light ray trajectory, where the distance of closest approach and the impact parameter are chosen so that the motion takes place outside the radius of certain circular orbits. Moreover, the strong gravitational lensing analysis performed here may provide a valuable tool for the search of cosmic strings. This motivates us to study in detail the gravitational lensing in the weak and strong deflection regimes due to a pair of causally disconnected black holes experiencing an acceleration in opposite directions under the action of forces generated by conical singularities.

The outline of this paper is as follows. In Section II, we derive the effective potential for a massless particle in the C-metric, we classify its critical points and we also derive an analytical formula for the radius of the circular orbits taking place on a family of invariant cones whose opening angles are also described by an exact formula. In Section III, we perform the Jacobi (in)stability analysis of the circular orbits. Section IV contains the construction of a Carter tetrad for the C-metric and the realization of an affine parametrization by means of a rotation of class III such that the Sachs optical scalar describing the shear effect on the light beam due to the gravitational field will coincide with a certain spin coefficient. These results are of fundamental importance  in Section V, where we compute the optical scalars in order to get insight into the contraction/expansion, rotation,  and/or  shearing experienced by a light beam travelling in a C-manifold. In particular, we show that the compression factor is the same as the one obtained in the Schwarzschild case. Hence, images of sources lying on the equatorial plane would experience the same degree of compression no matter if the lens is represented by a Schwarzschild black hole or by a black hole pulled by a cosmic string. We also show how away from the equatorial plane it may be possible to distinguish a Schwarzschild black hole from a C-black hole by probing into effects in the optical scalar $\theta$. Sec. VI deals with the gravitational lensing in the weak and strong regimes. In particular, the corresponding deflection angles are analytically computed when the light propagation occurs on a certain family of invariant cones and  their dependence on the black hole acceleration parameter is shown. Furthermore, our formulae correctly reproduce the corresponding ones in the Schwarzschild case in the limit of a vanishing acceleration parameter.

\section{Geodesic equations}\label{RKP}
The relativistic Kepler problem for the C-metric is defined by the geodesic equations \cite{FL}
\begin{equation}
\frac{d^2 x^\eta}{d\lambda^2}= -\Gamma^\eta_{\mu\nu}
\frac{dx^\mu}{d\lambda}\frac{dx^\nu}{d\lambda},\quad
\Gamma^\eta_{\mu\nu}=\frac{1}{2}g^{\eta\tau}\left(\partial_\mu g_{\tau\nu}+\partial_\nu g_{\tau\mu}-\partial_\tau g_{\mu\nu}\right)
\end{equation}
together with the subsidiary condition
\begin{equation}\label{constraint}
g_{\mu\nu} \frac{dx^\mu}{d\lambda}\frac{dx^\nu}{d\lambda}=-\epsilon,
\end{equation}
where $\epsilon=1$ , and $\epsilon=0$ corresponds to time-like and light-like geodesics, respectively, and $\lambda$ is an affine parameter. After the computation of the Christoffel symbols $\Gamma^\eta_{\mu\nu}$, we find that the equations of motion are represented by the following system of ordinary differential equations
\begin{eqnarray}
\frac{d^2 t}{d\lambda^2}&=&-\frac{\partial_r B}{B}\frac{dt}{d\lambda}\frac{dr}{d\lambda}-\frac{\partial_\vartheta B}{B}\frac{dt}{d\lambda}\frac{d\vartheta}{d\lambda},\label{g1}\\
\frac{d^2 r}{d\lambda^2}&=&-\frac{\partial_r B}{2A}\left(\frac{dt}{d\lambda}\right)^2
-\frac{\partial_r A}{2A}\left(\frac{dr}{d\lambda}\right)^2-\frac{\partial_\vartheta A}{A}\frac{dr}{d\lambda}\frac{d\vartheta}{d\lambda}
+\frac{\partial_r C}{2A}\left(\frac{d\vartheta}{d\lambda}\right)^2
+ \frac{\partial_r D}{2A}\left(\frac{d\phi}{d\lambda}\right)^2,\label{g2}\\
\frac{d^2\vartheta}{d\lambda^2}&=&-\frac{\partial_\vartheta B}{2C}\left(\frac{dt}{d\lambda}\right)^2
+\frac{\partial_\vartheta A}{2C}\left(\frac{dr}{d\lambda}\right)^2-\frac{\partial_r C}{C}\frac{dr}{d\lambda}\frac{d\vartheta}{d\lambda}
-\frac{\partial_\vartheta C}{2C}\left(\frac{d\vartheta}{d\lambda}\right)^2
+ \frac{\partial_\vartheta D}{2C}\left(\frac{d\phi}{d\lambda}\right)^2,\label{g3}\\
\frac{d^2\phi}{d\lambda^2}&=&-\frac{\partial_r D}{D}\frac{dr}{d\lambda}\frac{d\phi}{d\lambda}-\frac{\partial_\vartheta D}{D}\frac{d\vartheta}{d\lambda}\frac{d\phi}{d\lambda}   \label{g4}
\end{eqnarray}
with functions $A$, $B$, $C$, and $D$ defined in (\ref{ABCD}). Equation (\ref{g1}) can be rewritten as
\begin{equation}
\frac{1}{B}\frac{d}{d\lambda}\left(B\frac{dt}{d\lambda}\right)=0
\end{equation}
which integrated leads to
\begin{equation}\label{g5}
\frac{dt}{d\lambda}=\frac{\mathcal{E}}{B}.
\end{equation}
Since $B$ is dimensionless and in geometric units $dt/d\lambda$ is also dimensionless, it follows that the integration constant $\mathcal{E}$ is dimensionless. Furthermore, the C-metric admits a Killing vector $\partial/\partial t$ \cite{Lim}, and therefore, the aforementioned constant of motion must be related to the energy of the particle. More precisely, $\mathcal{E}$ represents the energy per unit mass \cite{FL} and hence, it is dimensionless in geometric units. Similarly, we can integrate equation (\ref{g4}) to obtain
\begin{equation}\label{g6}
\frac{d\phi}{d\lambda}=\frac{\ell}{D}.
\end{equation}
Note that the C-metric admits the Killing vector $\partial/\partial\phi$ \cite{Lim}, and therefore, the integration constant $\ell$ can be related to the angular momentum of the particle. More precisely, $\ell$ is the angular momentum per unit mass \cite{FL}, and in geometric units, it has dimension of length. We point out that the second equation appearing in $(14)$ in \cite{Lim} and corresponding to our (\ref{g6}) should have a factor $C_0^2$ in the denominator. If we substitute (\ref{g5}) and (\ref{g6}) into (\ref{g2}) and (\ref{g3}), we get  
\begin{eqnarray}
\frac{d^2 r}{d\lambda^2}&=&-\frac{\partial_r A}{2A}\left(\frac{dr}{d\lambda}\right)^2-\frac{\partial_\vartheta A}{A}\frac{dr}{d\lambda}\frac{d\vartheta}{d\lambda}
+\frac{\partial_r C}{2A}\left(\frac{d\vartheta}{d\lambda}\right)^2
-\frac{\mathcal{E}^2}{2}\frac{\partial_r B}{AB^2}+\frac{\ell^2}{2}\frac{\partial_r D}{AD^2},\label{g7}\\
\frac{d^2\vartheta}{d\lambda^2}&=&-\frac{\partial_\vartheta C}{2C}\left(\frac{d\vartheta}{d\lambda}\right)^2-\frac{\partial_r C}{C}\frac{dr}{d\lambda}\frac{d\vartheta}{d\lambda}+\frac{\partial_\vartheta A}{2C}\left(\frac{dr}{d\lambda}\right)^2-\frac{\mathcal{E}^2}{2}\frac{\partial_\vartheta B}{CB^2}+\frac{\ell^2}{2}\frac{\partial_\vartheta D}{CD^2}.\label{g8}
\end{eqnarray}
At this point a couple of remarks are in order. After a careful scrutiny of equation (16) in \cite{Lim}, we noticed there that the corresponding coefficient multiplying the term $\dot{r}^2$ has a spurious extra factor $2$. Furthermore, the coefficient going together with the term $\dot{r}\dot{\vartheta}$ should have $P$ replaced by $r$. Note that the presence of $P$ in the denominator of the term containing $\dot{r}\dot{\vartheta}$ in (16) would cause this term to have different dimension as all other terms in (16). The expressions for the coefficient functions entering in (\ref{g7}) and (\ref{g8}) can be found in the Appendix. Differently as in the Schwarzschild case, due to the lack of spherical symmetry equation (\ref{g8}) cannot be solved without loss of generality by assuming that the motion of the particle takes place on the equatorial plane, i.e. $\vartheta=\pi/2$. Furthermore, the constraint equation (\ref{constraint}) can be expressed with the help of (\ref{g5}) and (\ref{g6}) as
\begin{equation}\label{epf1}
A\left(\frac{dr}{d\lambda}\right)^2+C\left(\frac{d\vartheta}{d\lambda}\right)^2=-\epsilon+\frac{\mathcal{E}^2}{B}-\frac{\ell^2}{D}.
\end{equation}
If we multiply (\ref{epf1}) by a factor $1/2$ so that in the limit of vanishing $\alpha$ equation (\ref{epf1}) reproduces correctly equation (25.26) in \cite{FL} for the Schwarzschild case, and introduce the effective potential
\begin{equation}
U_{eff}(r,\vartheta)=\frac{1}{2}\left(\epsilon B+\frac{\ell^2 B}{D}\right),
\end{equation}
we can rewrite equation (\ref{epf1}) in a more suitable form to study the motion of a particle in the C-metric, namely
\begin{equation}\label{cos}
\frac{AB}{2}\left(\frac{dr}{d\lambda}\right)^2+\frac{BC}{2}\left(\frac{d\vartheta}{d\lambda}\right)^2+U_{eff}=E,
\end{equation}
where we set $E=\mathcal{E}^2/2$ and
\begin{equation}
AB=F^2,\quad BC=r^2F^2\frac{f}{g}.
\end{equation}
Notice that $A$, $B$ and $C$ are positive definite and therefore $E-U_{eff} \ge 0$ as in classical mechanics. The equality, $E=U_{eff}$, corresponds to a circular orbit and a critical point of the effective potential. Before proceeding to study light bending in the C-metric, we recall that in the case of null geodesics we have $\epsilon=0$ and therefore, the effective potential takes on the more simpler form
\begin{equation}\label{Ueff}
\mathfrak{V}(r,\vartheta)=\frac{\ell^2 B}{2D}.
\end{equation}
At this point a couple of remarks are in order. First of all, the effective potential $V_{eff}$ introduced in \cite{Lim} in equation (19) is related to our potential as follows
\begin{equation}
2U_{eff}(r,\vartheta)=V^2_{eff}(r,\vartheta).
\end{equation}
It is straightforward to verify that the effective potential $V_{eff}=\sqrt{2U_{eff}}$ defined by \cite{Lim} does not reproduce in the limit $\alpha\to 0^{+}$ and for $\vartheta=\pi/2$ the effective potential for a particle in the Schwarzschild metric which is given in geometric units by \cite{FL,CH,CAR}
\begin{equation}\label{Schw_m}
V_{eff,Sch}(r)=\left\{
\begin{array}{cc}
-\frac{M}{r}+\frac{\ell^2}{2r^2}-\frac{M\ell^2}{r^3}, & m\neq 0\\
\frac{\ell^2}{2r^2}-\frac{M\ell^2}{r^3}, & m=0
\end{array}
\right.,
\end{equation}
where $m$ denotes the mass of the particle. Furthermore, in the Schwarzschild case and for $\epsilon=0$, the critical point of the effective potential coincides with the critical point of the corresponding geodesic equation (see Appendix~\ref{AppS}). We will see that the same continues to be true also in the case of the C-metric. We conclude this part by expressing the geodesic equations (\ref{g7}) and (\ref{g8}) subject to the constraint (\ref{cos}) in a form that simplifies considerably the analysis of the critical point(s) arising from the dynamical system associated to the new system of equations and the study of the Jacobian (in)stability of the circular orbits. To this purpose, we replace (\ref{cos}) into (\ref{g7}) and (\ref{g8}) to obtain
\begin{eqnarray}
&&\frac{d^2 r}{d\lambda^2}+[\partial_r\ln{\sqrt{AC}}]\left(\frac{dr}{d\lambda}\right)^{2}+(\partial_\vartheta\ln{A}) \frac{dr}{d\lambda}\frac{d\vartheta}{d\lambda}+\frac{E}{AB}\partial_r\ln{\frac{B}{C}}=0,\label{din1}\\
&&\frac{d^2 \vartheta}{d\lambda^2}+[\partial_\vartheta\ln{\sqrt{AC}}]\left(\frac{d\vartheta}{d\lambda}\right)^{2}+(\partial_r\ln{C})\frac{dr}{d\lambda}\frac{d\vartheta}{d\lambda}+\frac{\ell^2}{2CD}\partial_\vartheta\ln{\frac{A}{D}}=0.\label{din2}
\end{eqnarray}
Let us investigate the critical points of the effective potential (\ref{Ueff}). First of all, the condition $\partial_r\mathfrak{V}=0$ gives rise to the quadratic equation
\begin{equation}\label{quad}
\alpha^2 Mr^2+r-3M=0.
\end{equation}
It is comforting to see that in the limit of vanishing $\alpha$, equation (\ref{quad}) has two coinciding roots reproducing the radius of the photon sphere $r_\gamma=3M$ for a Schwarzschild black hole. Since the radial variable $r$ belongs to the interval $(2M,1/\alpha)$, the only acceptable root for (\ref{quad}) is
\begin{equation}\label{crit_point_r}
r_c=\frac{6M}{1+\sqrt{1+12\alpha^2 M^2}}.
\end{equation}
\begin{figure}\label{det}
\includegraphics[scale=0.35]{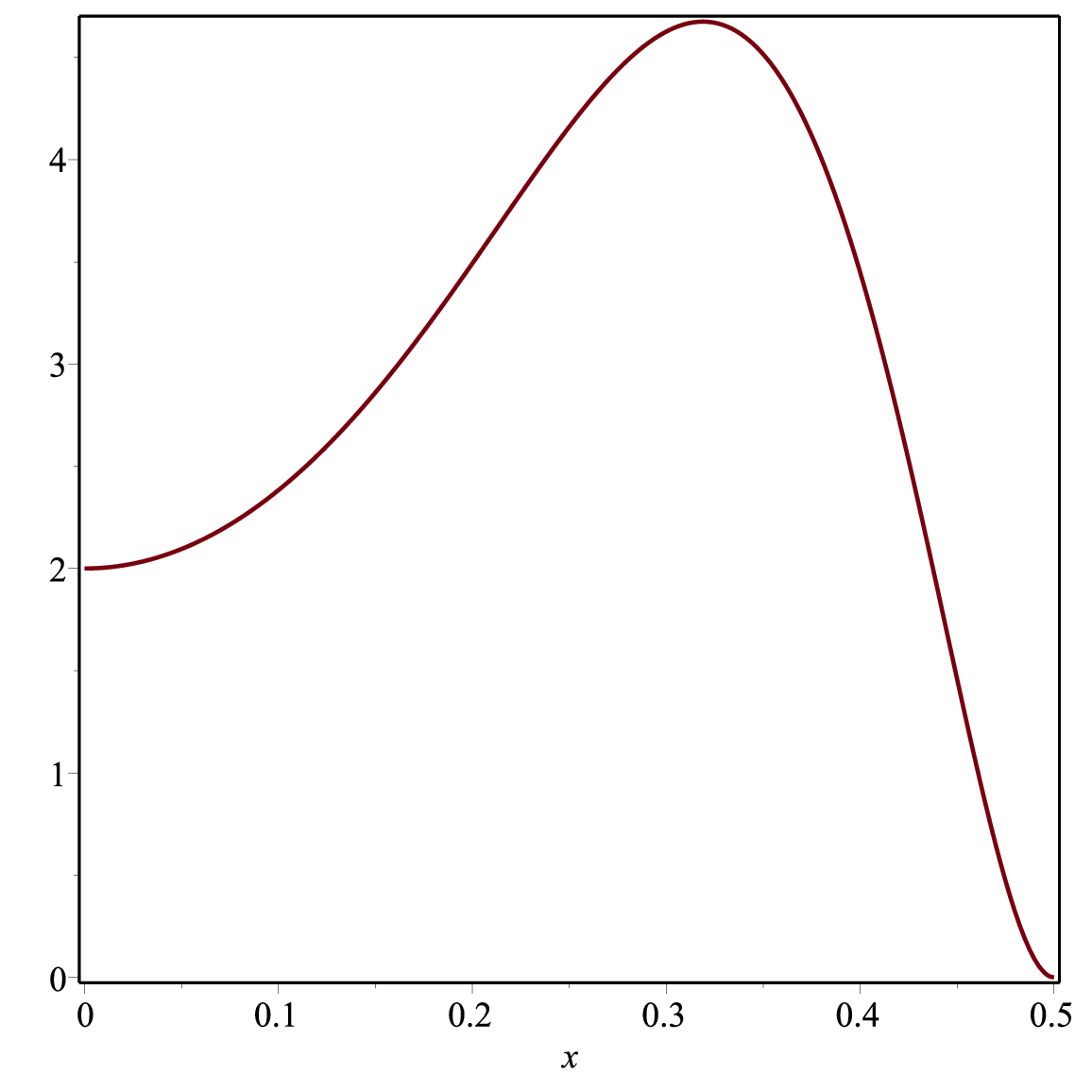}
\caption{\label{fig0}
Plot of the functions $S$ defined in (\ref{S}) on the interval $0<x<\frac{1}{2}$.}
\end{figure}
From (\ref{crit_point_r}), we immediately see that $r_c<r_\gamma$. This implies that a null circular orbit for a photon in the C-metric takes place at a closer distance to the event horizon than in the Schwarzschild case. Furthermore, if we require that $\partial_\vartheta\mathfrak{V}=0$, we end up with the equation
\begin{equation}\label{thetaeq}
3\alpha M\cos^2{\vartheta}+\cos{\vartheta}-\alpha M=0.
\end{equation}
The root
\begin{equation}
\cos{\vartheta_1}=-\frac{1+\sqrt{1+12\alpha^2 M^2}}{6\alpha M}=\frac{2\alpha M}{1-\sqrt{1+12\alpha^2 M^2}}
\end{equation}
must be disregarded because we would have $\cos{\vartheta_1}<-1$ for $\alpha M\in(0,1/2)$. Hence, the only admissible solution to (\ref{thetaeq}) is  
\begin{equation}\label{crit_point_theta}
\vartheta_c=\arccos{\left(\frac{2\alpha M}{1+\sqrt{1+12\alpha^2 M^2}}\right)}.
\end{equation}
Since $\alpha M\in(0,1/2)$ and $\vartheta_c$ is monotonically decreasing in the variable $\alpha M$, it is straightforward to verify that the angular variable associated to the critical points cannot take on any value from $0$ to $\pi$ but it is constrained to the range $(\vartheta_{c,min},\pi/2)$ with $\vartheta_{c,min}=\arccos{(1/3)}\approx 70.52^{\circ}$. Concerning the classification of these critical points, if we compute the determinant $\Delta$ of the Hessian matrix associated to our effective potential and we evaluate it at the critical point  $(r_c,\vartheta_c)$ found above, it turns out that
\begin{equation}\label{Hessian}
\Delta(r_c,\vartheta_c)=-\frac{\ell^4}{139968 M^6\kappa^4}\frac{(1+\tau)^9}{(1+\tau+4x^2)^3}\frac{S(x)}{T(x)}
\end{equation}
with $x:=\alpha M$, $\tau=\tau(x):=\sqrt{1+12x^2}$ and
\begin{eqnarray}
S(x)&:=&(1728\tau+8640)x^{10}+(1008\tau-720)x^8-(492\tau+828)x^6+(25-35\tau)x^4+(13\tau+19)x^2+1+\tau,\label{S}\\
T(x)&:=&32 x^8+(32\tau+176)x^6+(48\tau+114)x^4+(14\tau+20)x^2+1+\tau.
\end{eqnarray}
Since $x\in(0,1/2)$ the term $T(x)$ in (\ref{Hessian}) is always positive and therefore, the sign of (\ref{Hessian}) is controlled by term $S(x)$. As it can be evinced from Fig.~\ref{fig0}, the function $S(x)$ is always positive on the interval $(0,1/2)$, and therefore, $\Delta(r_c,\vartheta_c)<0$. This means that the critical point $(r_c,\vartheta_c)$ of the effective potential is a saddle point. As a check of the validity of our computation, we computed the value of the effective potential at the saddle point, namely
\begin{equation}\label{V_at_sp}
\mathfrak{V}(r_c,\vartheta_c)=\frac{\ell^2[216 x^6-54x^2+\tau(1-6x^2)+1]}{27M^2\kappa^2(4x^2+\tau+1)},\quad
x=\alpha M,\quad \tau=\tau(x)=\sqrt{1+12x^2},
\end{equation}
and we verified that in the limit of vanishing $\alpha$, it reproduces correctly the value $\ell^2/(54M^2)$ obtained from (25.27) in \cite{FL} for the maximum of the effective potential computed at the radius of the photon sphere. Furthermore, (\ref{V_at_sp}) attains its maximum for $\alpha=0$ (Schwarzschild case), and then, it decreases monotonically to zero as $\alpha M$ approaches the value $1/2$. As in the Schwarzschild case (see Appendix~\ref{AppS}), the critical point of the effective potential for massless particles in the C-metric coincides with the critical point of the dynamical system represented by (\ref{din1}) and (\ref{din2}). To see that, let $r=r_0$ and $\vartheta=\vartheta_0$ be the coordinates of a candidate critical point for the system (\ref{din1}) and (\ref{din2}). Then, the latter  simplifies to the following system of equations
\begin{equation}
\left.\partial_r\left(\frac{B}{C}\right)\right|_{(r_0,\vartheta_0)}=0,\quad
\left.\partial_\vartheta\left(\frac{A}{D}\right)\right|_{(r_0,\vartheta_0)}=0.
\end{equation}
By means of (\ref{ABCD}) the above equations reduce to
\begin{equation}\label{radius}
\left.\frac{d}{dr}\left(\frac{f(r)}{r^2}\right)\right|_{r=r_0}=0,\quad
\left.\frac{d}{d\vartheta}\left(\frac{1}{g\sin^2{\vartheta}}\right)\right|_{\vartheta=\vartheta_0}=0.
\end{equation}
At this point, a straightforward computation shows that the first equation in (\ref{radius}) reduces to the quadratic equation (\ref{quad}) and therefore, we conclude that $r_0=r_c$ with $r_c$ given by (\ref{crit_point_r}). Concerning the second equation in (\ref{radius}), it can be easily seen that its r.h.s is equivalent to 
\begin{equation}
\frac{d}{d\vartheta}\left(\frac{1}{g\sin^2{\vartheta}}\right)=-\frac{2(3\alpha M\cos^2{\vartheta}+\cos{\vartheta}-\alpha M)}{\sin^2{\vartheta}(1+2\alpha M\cos{\vartheta})^2}
\end{equation}
and hence, it will vanish when (\ref{thetaeq}) is satisfied. At this point, we can perform the same analysis done to study the roots of (\ref{thetaeq}) and it follows that $\vartheta_0=\vartheta_c$.

\section{Jacobi stability analysis of the circular orbits}
In this section, we dwell with the problem of determining whether or not the class of circular orbits we previously found are Jacobi stable. To study the Jacobi (in)stability of photon circular orbits with radius $r_c$ and occurring on a cone with opening angle  $\vartheta=\vartheta_c$, we shall use the KCC theory which represents a powerful mathematical method for the analysis of dynamical systems \cite{21,22,23,28,Bom}. To this purpose, we consider null particles and rewrite (\ref{din1}) and (\ref{din2}) as a dynamical system of the form 
\begin{equation}\label{sist}
\frac{d^2 x^i}{d\lambda^2}+g^i(x^1,x^2,y^1,y^2)=0
\end{equation}
where
\begin{eqnarray}
g^1(x^1,x^2,y^1,y^2)&=&[\partial_1\ln{\sqrt{AC}}](y^1)\strut^{2}+(\partial_2\ln{A}) y^1y^2+\frac{E}{AB}\partial_1\ln{\frac{B}{C}},\label{gg1}\\
g^2(x^1,x^2,y^1,y^2)&=&[\partial_2\ln{\sqrt{AC}}](y^2)\strut^{2}+(\partial_1\ln{C}) y^1y^2+\frac{\ell^2}{2CD}\partial_2\ln{\frac{A}{D}}\label{gg2}
\end{eqnarray}
with $x^1:=r$, $x^2:=\vartheta$, and $y^i=dx^{i}/d\lambda$ for $i=1,2$. Furthermore, we make the reasonable assumption that $g^1$ and $g^2$ are smooth functions in a neighbourhood of the initial condition $(x^1_0,x^2_0,y^1_0,y^2_0,\lambda_c)=(r_c,\vartheta_c,0,0,\lambda_c)\in\mathbb{R}^5$. If we perturb the geodesic trajectories of the system (\ref{sist}) into neighbouring ones according to 
\begin{equation}\label{C}
\widehat{x}^i(\lambda)=x^i(\lambda)+\eta\xi^i(\lambda),
\end{equation}
where $|\eta|$ is a small parameter, and $\xi^i(\lambda)$ represents the components of some contravariant vector field defined along the geodesic trajectory $x^i(\lambda)$, the equation governing the perturbative part in (\ref{C}) can be obtained by replacing first (\ref{C}) into (\ref{sist}) and by letting then $\eta\to 0$. This procedure leads to the equation
\begin{equation}\label{xi}
\frac{d^2\xi^i}{d\lambda^2}+2N^i_r\frac{d\xi^r}{d\lambda}+\frac{\partial g^i}{\partial x^r}\xi^r=0,
\end{equation}
where
\begin{equation}\label{enne}
N^i_j=\frac{1}{2}\frac{\partial g^i}{\partial y^j}
\end{equation}
defines the coefficients of a non-linear connection $N$ on the tangent bundle. The latter also enters the definition of the KCC covariant differential, namely
\begin{equation}\label{KCCdiff}
\frac{D\sigma^i}{d\lambda}=\frac{d\sigma^i}{d\lambda}+N^i_j\sigma^j
\end{equation}
with $\sigma=\sigma^i\partial/\partial x^i$ some contravariant vector field. Furthermore, equation (\ref{xi}) can be written in terms of (\ref{KCCdiff}) in the covariant form
\begin{equation}\label{Jacobi}
\frac{D^2\xi^i}{d\lambda^2}=P^i_j\xi^j.
\end{equation}
Equation (\ref{Jacobi}) is called the Jacobi equation or the variation equation associated to the system (\ref{sist}). Proceeding as in \cite{Bom} five geometrical invariants can be obtained for the system (\ref{sist}). However, only the second invariant controls the Jacobi (in)stability of the system, namely the tensor
\begin{equation}\label{KKC2}
P^i_j=-\frac{\partial g^i}{\partial x^j}-g^rG^i{}_{rj}+y^r\frac{\partial N^i_j}{\partial x^r}+N^i_r N^r_j+\frac{\partial N^i_j}{\partial\lambda},\quad G^i{}_{rj}=\frac{\partial N^i_r}{\partial y^j}
\end{equation}
where $G^i{}_{rj}$ is called the Berwald connection \cite{Anto,Miron}. Note that the term $\partial N^i_j/\partial\lambda$ in (\ref{KKC2}) will not contribute because the system (\ref{sist}) is autonomous in the variable $\lambda$. More precisely, if $\xi^i=v(\lambda)\nu^i$ is a Jacobi field with speed $v$ along the geodesic $x^i(\lambda)$ where $\nu^i$ is the unit normal vector field, then the Jacobi field equation (\ref{Jacobi}) can be represented in the scalar form as \cite{Bao} 
\begin{equation}\label{jac2}
\frac{d^2 v}{d\lambda^2}+Kv=0
\end{equation}
with $K$ denoting the flag curvature of the manifold. The sign of $K$ governs the geodesic trajectories, that is if $K<0$, then such trajectories disperse, i.e. they are Jacobi unstable, and otherwise, if $K>0$, they tend to focus together, i.e. they are Jacobi stable. Instead of studying the Jacobi (in)stability of the circular orbits by analyzing the sign of $K$, it turns out that it is more convenient to use the following result: {\it{an integral curve $\gamma$ of (\ref{sist}) is Jacobi stable if and only if the real parts of the eigenvalues of the second KCC invariant $P^i_j$ are strictly negative everywhere along $\gamma$, and Jacobi unstable otherwise}}. We refer to \cite{1,6,Bom} for the proof of this statement. In order to apply this result, let us introduce the matrix
\begin{equation}
P:=\left(
\begin{array}{cc}
P^1_1 & P^1_2\\
P^2_1 & P^2_2
\end{array}
\right)
\end{equation}
evaluated at a circular orbit with $x^1=r_c$ and $x^2=\vartheta_c$ given by (\ref{crit_point_r}) and (\ref{crit_point_theta}), respectively. Then, the associated characteristic equation is
\begin{equation}\label{cee}
\mbox{det}\left(
\begin{array}{cc}
P^1_1(r_c,\vartheta_c)-\lambda & P^1_2(r_c,\vartheta_c)\\
P^2_1(r_c,\vartheta_c) & P^2_2(r_c,\vartheta_c)-\lambda
\end{array}
\right)=0.
\end{equation}
The computation of the entries of the matrix $P$ can be optimized if we observe that the connection $N$ vanish along a circular orbit with $x^1=r_c$ and $x^2=\vartheta_c$. This is due to the fact that the functions $g^i$ given by (\ref{gg1}) and (\ref{gg2}) are quadratic in the variables $y^1$ and $y^2$, and hence, the terms $N^i_j$ defined through (\ref{enne}) contain only linear combinations of $y^1$ and $y^2$ but  $y^i=dx^i/d\lambda=0$ along a circular orbit with $x^1=r_c$ and $x^2=\vartheta_0$. Similarly, terms of the form $y^r(\partial N^i_j/\partial x^r)$ must also vanish when evaluated along the circular trajectories since $y^i=0$ there. Hence, the only terms contributing to the computation of the eigenvalues are the first two terms on the l.h.s. of (\ref{KKC2}). Let $x=\alpha M$ and $\rho=r/M$. After a lengthy but straightforward computation where we made use of (\ref{radius}) we find that
\begin{eqnarray}
P^1_1(\rho_c,\vartheta_c)&=&\left.-\frac{E}{M^2}\left[\frac{\rho^2}{fF^2}\frac{d^2}{d\rho^2}\left(\frac{f}{\rho^2}\right)\right] \right|_{(\rho_c,\vartheta_c)},\\
P^2_2(\rho_c,\vartheta_c)&=&-\left.\left(\frac{L}{M}\right)^2\left[\frac{g}{2\kappa^2\rho^4 F^2}\frac{d^2}{d\vartheta^2}\left(\frac{1}{g\sin^2{\vartheta}}\right) \right]\right|_{(\rho_c,\vartheta_c)},\\
P^1_2(\rho_c,\vartheta_c)&=&\left.\frac{E}{M}\left[\frac{3\rho^2}{2fF^3}\left(\partial_\vartheta F\right)\frac{d}{d\rho}\left(\frac{f}{\rho^2}\right)\right]\right|_{(\rho_c,\vartheta_c)}=0,\label{beauty}\\
P^2_1(\rho_c,\vartheta_c)&=&-\frac{L^2}{M^3}\left.\left\{\frac{1}{2\kappa^2}\left[\partial_\rho\left(\frac{g}{\rho^4 F^2}\right)+\frac{g}{2\rho^6 F^3}\partial_\rho\left(\rho^2 F\right)\right]\frac{d}{d\vartheta}\left(\frac{1}{g\sin^2{\vartheta}}\right)\right\}\right|_{(\rho_c,\vartheta_c)}=0
\end{eqnarray}
with $L:=\ell/M$. This implies that the zeroes of the characteristic equation (\ref{cee}) are real and given by
\begin{equation}\label{eigenwerte}
\lambda_1=-\frac{E}{M^2}\Omega_{1}(\rho_c,\vartheta_c),\quad
\lambda_2=-\frac{L^2}{M^2}\Omega_{2}(\rho_c,\vartheta_c),
\end{equation}
where 
\begin{equation}\label{omegas}
\Omega_{1}(\rho_c,\vartheta_c)=\left.\left[\frac{\rho^2}{fF^2}\frac{d^2}{d\rho^2}\left(\frac{f}{\rho^2}\right)\right]\right|_{(\rho_c,\vartheta_0)},\quad
\Omega_{2}(\rho_c,\vartheta_c)=\left.\left[\frac{g}{2\kappa^2\rho^4 F^2}\frac{d^2}{d\vartheta^2}\left(\frac{1}{g\sin^2{\vartheta}}\right)\right]\right|_{(\rho_c,\vartheta_0)}.
\end{equation}
\begin{figure}\label{onemore}
\includegraphics[scale=0.35]{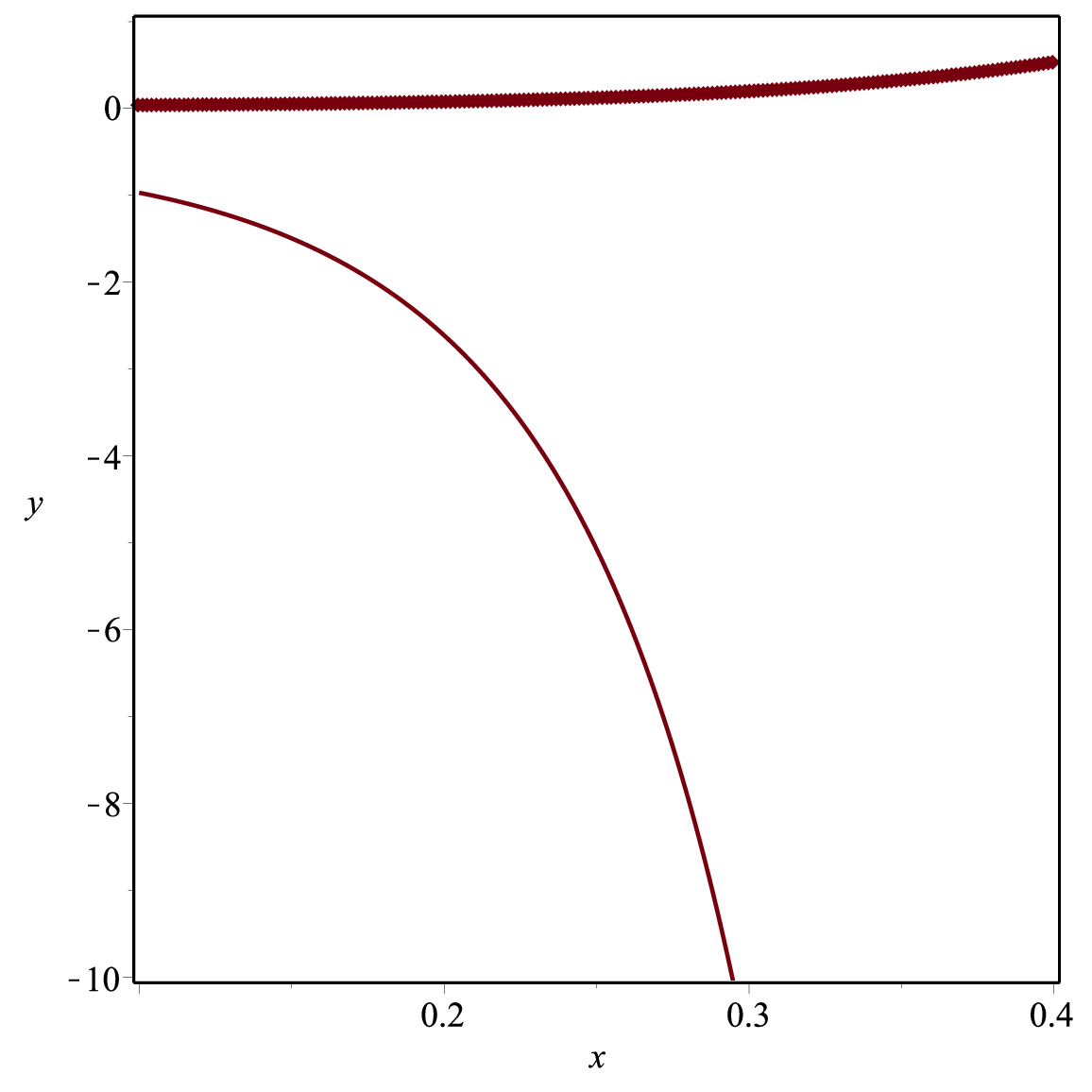}
\caption{\label{fig04}
Plot of the functions $\Omega_{1}$ (thin solid line) and $\Omega_2$ (thick solid line) defined in (\ref{omegas}) as a functions of $x=\alpha M$ on the interval $(0,1/2)$. The function $\Omega_{1}$  is always negative on $(0,1/2)$. }
\end{figure}
From Fig.~\ref{fig04}, we immediately see that $\Omega_{1}$ is always negative for $x\in(0,1/2)$ and therefore, the eigenvalue $\lambda_1$ is always strictly positive. As a consequence, photon circular orbits with radius $r_c$ given by (\ref{crit_point_r}) occurring on the cones $\vartheta=\vartheta_c$ with $\vartheta_0\in(\pi/2,\pi)$ given by (\ref{crit_point_theta}) are Jacobi unstable.

\section{Carter tetrad for the C-metric}
We construct a special tetrad for the C-metric which has the advantage to considerably simplify the computation of the Sachs optical scalars done in the next section. Following the procedure outlined in \cite{Carter} (see equation (5.119) therein), we introduce the differential forms
\begin{equation}
\theta^1_\mu=\sqrt{\frac{F}{f}}~dr,\quad
\theta^2_\mu=r\sqrt{\frac{F}{g}}~d\vartheta,\quad
\theta^3_\mu=\kappa r\sqrt{Fg}\sin{\vartheta}~d\varphi,\quad
\theta^4_\mu=\sqrt{fF}~dt
\end{equation}
and construct a symmetric null tetrad $(\bm{\ell},\mathbf{n},\mathbf{m},\overline{\mathbf{m}})$ with
\begin{equation}
\ell_\mu=\frac{\theta^1_\mu+\theta^4_\mu}{\sqrt{2}},\quad
n_\mu=\frac{\theta^1_\mu-\theta^4_\mu}{\sqrt{2}},\quad
m_\mu=\frac{\theta^2_\mu+i\theta^3_\mu}{\sqrt{2}},\quad
\overline{m}_\mu=\frac{\theta^2_\mu-i\theta^3_\mu}{\sqrt{2}}.
\end{equation}
such that the normalization and orthogonality relations 
\begin{equation}\label{tetrad_relations}
\bm{\ell}\cdot\mathbf{n}=1,\quad 
\mathbf{m}\cdot\overline{\mathbf{m}}=-1,\quad
\bm{\ell}\cdot\mathbf{m}=\bm{\ell}\cdot\overline{\mathbf{m}}=\mathbf{n}\cdot\mathbf{m}=\mathbf{n}\cdot\overline{\mathbf{m}}=0
\end{equation}
are satisfied. Furthermore, the spin coefficients are given by \cite{CH}
\begin{eqnarray}
\kappa&=&\gamma_{(3)(1)(1)},\quad
\sigma=\gamma_{(3)(1)(3)},\quad
\lambda=\gamma_{(2)(4)(4)},\quad
\nu=\gamma_{(2)(4)(2)},\quad
\rho=\gamma_{(3)(1)(4)},\quad
\mu=\gamma_{(2)(4)(3)},\label{mu}\\
\tau&=&\gamma_{(3)(1)(2)},\quad
\pi=\gamma_{(2)(4)(1)},\quad
\epsilon=\frac{1}{2}\left[\gamma_{(2)(1)(1)}+\gamma_{(3)(4)(1)}\right],\quad
\gamma=\frac{1}{2}\left[\gamma_{(2)(1)(2)}+\gamma_{(3)(4)(2)}\right],\label{gamma}\\
\alpha&=&\frac{1}{2}\left[\gamma_{(2)(1)(4)}+\gamma_{(3)(4)(4)}\right],\quad
\beta=\frac{1}{2}\left[\gamma_{(2)(1)(3)}+\gamma_{(3)(4)(3)}\right]\label{beta}
\end{eqnarray}
with $(a)$ denoting the tetrad index and $a=1,\cdots,4$. The Ricci rotation coefficients are expressed as
\begin{equation}\label{Ricci}
\gamma_{(a)(b)(c)}=\frac{1}{2}\left[\lambda_{(a)(b)(c)}+\lambda_{(c)(a)(b)}-\lambda_{(b)(c)(a)}\right],\quad
\lambda_{(a)(b)(c)}=e_{(b)i,j}\left[e_{(a)}{}^ie_{(c)}{}^j-e_{(a)}{}^je_{(c)}{}^i\right], 
\end{equation}
where
\begin{equation}
(e_{(a)}{}^i)=\left(
\begin{array}{cccc}
\ell^1 & \ell^2 & \ell^3 & \ell^4\\
n^1    & n^2    & n^3    & n^4 \\
m^1    & m^2    & m^3    & m^4 \\
\overline{m^1}    & \overline{m^2}    & \overline{m^3}    & \overline{m^4}
\end{array}
\right)=\left(
\begin{array}{cccc}
\frac{1}{\sqrt{2fF}} & -\sqrt{\frac{f}{2F}} & 0 & 0\\
\frac{1}{\sqrt{2fF}} &  \sqrt{\frac{f}{2F}} & 0 & 0\\
0 & 0 & -\frac{1}{r}\sqrt{\frac{g}{2F}} & -\frac{i}{\kappa r\sin{\vartheta}\sqrt{2Fg}} \\
0 & 0 & -\frac{1}{r}\sqrt{\frac{g}{2F}} & \frac{i}{\kappa r\sin{\vartheta}\sqrt{2Fg}}
\end{array}
\right)
\end{equation}
and its inverse is denoted by $(e^{(b)}{}_i)$. We recall that tetrad indices label rows and tensor indices label columns. Taking into account that $e_{(a)i}=g_{ik}e_{(a)}{}^k$ and the metric tensor of the C-metric is represented by a diagonal matrix, we find that
\begin{equation}
(e_{(a)i})=\left(
\begin{array}{cccc}
\ell_1 & \ell_2 & \ell_3 & \ell_4\\
n_1    & n_2    & n_3    & n_4 \\
m_1    & m_2    & m_3    & m_4 \\
\overline{m_1}    & \overline{m_2}    & \overline{m_3}    & \overline{m_4}
\end{array}
\right)=\left(
\begin{array}{cccc}
\sqrt{\frac{fF}{2}} & \sqrt{\frac{F}{2f}} & 0 & 0\\
\sqrt{\frac{fF}{2}} & -\sqrt{\frac{F}{2f}} & 0 & 0\\
0 & 0 & r\sqrt{\frac{F}{2g}} & i\kappa r\sin{\vartheta}\sqrt{\frac{Fg}{2}} \\
0 & 0 & r\sqrt{\frac{F}{2g}} &-i\kappa r\sin{\vartheta}\sqrt{\frac{Fg}{2}}
\end{array}
\right).
\end{equation}
The spin coefficients are computed to be
\begin{eqnarray}
\kappa&=&\sigma=\lambda=\nu=0,\quad
\rho=\mu=\frac{1}{rF}\sqrt{\frac{f}{2}}\frac{\partial (r\sqrt{F})}{\partial r},\quad
\tau=-\pi=\frac{1}{2rF}\sqrt{\frac{g}{2F}}\frac{\partial F}{\partial\vartheta},\\
\epsilon&=&\gamma=-\frac{1}{2\sqrt{2}F}\frac{\partial\sqrt{fF}}{\partial r},\quad
\beta=-\alpha=-\frac{1}{2\sqrt{2}Fr\sin{\vartheta}}\frac{\partial \sqrt{Fg}\sin{\vartheta}}{\partial\vartheta}.
\end{eqnarray}
At this point a remark is in order. Since the spin coefficient $\kappa$ vanishes, the $\bm{\ell}$-vectors form a congruence of null geodesics. This is not surprising because the C-metric is a Petrov type D metric. However, such null geodesics are not affinely parametrized because the spin coefficient $\epsilon$ does not vanish. The fact that $\kappa=0$ allows to construct an affine parametrization by means of a rotation of class III (see $\S$7(g) eq. (347) in \cite{CH}) to be treated later, such that it will not affect neither the direction of $\bm{\ell}$ nor the spin coefficient $\kappa$. As a nice side effect, the Sachs optical scalar describing the shear effect on the light beam due to the gravitational field will coincide with the spin coefficient $\sigma$. To this purpose, we transform the vectors $\bm{\ell}$ and $\bm{n}$ according to
\begin{equation}\label{trafo}
\ell_{new,i}=\frac{\ell_i}{\sqrt{fF}},\quad n_{new,i}=\sqrt{fF}n_i,\quad i=1,\cdots,4
\end{equation}
while keeping $\bm{m}$ and $\overline{\bm{m}}$ unchanged. It can easily checked that  the metric elements 
\begin{equation}
g_{ij}=\ell_{new,i}n_{new,j}+\ell_{new,j}n_{new,i}-m_i\overline{m}_j-m_j\overline{m}_i
\end{equation}
are invariant under the transformation (\ref{trafo}) and moreover, the tetrad 
\begin{equation}\label{new tetrad}
\ell_{new,i}=\left(\frac{1}{\sqrt{2}},\frac{1}{\sqrt{2}f},0,0\right),\quad
n_{new,i}=\left(\frac{f F}{\sqrt{2}},-\frac{F}{\sqrt{2}},0,0\right),\quad m_{new,i}=m_i,\quad \overline{m}_{new,i}=\overline{m}_i
\end{equation}
satisfies the nullity, orthogonality, and normalization conditions. From now on, we omit the "new" subscript in the tetrad (\ref{new tetrad}) and it is understood that the following results are obtained using (\ref{new tetrad}). To compute the spin coefficients, we rely again on (\ref{mu})-(\ref{Ricci}) with
\begin{equation}
(e_{(a)}{}^i)=\left(
\begin{array}{cccc}
\frac{1}{\sqrt{2}fF} & -\frac{1}{\sqrt{2}F} & 0 & 0\\
\frac{1}{\sqrt{2}} &  \frac{f}{\sqrt{2}} & 0 & 0\\
0 & 0 & -\frac{1}{r}\sqrt{\frac{g}{2F}} & -\frac{i}{\kappa r\sin{\vartheta}\sqrt{2Fg}} \\
0 & 0 & -\frac{1}{r}\sqrt{\frac{g}{2F}} & \frac{i}{\kappa r\sin{\vartheta}\sqrt{2Fg}}
\end{array}
\right),\quad
(e_{(a)i})=\left(
\begin{array}{cccc}
\frac{1}{\sqrt{2}} & \frac{1}{\sqrt{2}f} & 0 & 0\\
\frac{fF}{\sqrt{2}} & -\frac{F}{\sqrt{2}} & 0 & 0\\
0 & 0 & r\sqrt{\frac{F}{2g}} & i\kappa r\sin{\vartheta}\sqrt{\frac{Fg}{2}} \\
0 & 0 & r\sqrt{\frac{F}{2g}} &-i\kappa r\sin{\vartheta}\sqrt{\frac{Fg}{2}}
\end{array}
\right).
\end{equation}
In the new tetrad (\ref{new tetrad}), the spin coefficients are given by
\begin{eqnarray}
\kappa&=&\sigma=\lambda=\nu=\epsilon=0,\quad
\rho=\frac{1}{r\sqrt{2F}},\quad\mu=\frac{f}{r}\sqrt{\frac{F}{2}},\quad
\tau=-\pi=\frac{1}{2rF}\sqrt{\frac{g}{2F}}\frac{\partial F}{\partial\vartheta},\\
\gamma&=&-\frac{1}{2\sqrt{2}F}\frac{\partial(fF)}{\partial r},\quad
\beta=-\frac{1}{2\sqrt{2}rF\sin{\vartheta}}\frac{\partial \sqrt{Fg}\sin{\vartheta}}{\partial\vartheta}+\frac{\tau}{2},\quad
\alpha=\frac{1}{2\sqrt{2}rF\sin{\vartheta}}\frac{\partial \sqrt{Fg}\sin{\vartheta}}{\partial\vartheta}+\frac{\tau}{2}.\label{aa}
\end{eqnarray}
In the next section, we make use of the tetrad (\ref{new tetrad}) to compute the Sachs optical scalars.

\section{Sachs optical scalars}
The study of the Jacobi equation, i.e. the equation of geodesic deviation, represents the starting point in calculations aiming to determine distance measures, image distortion, and image brightness of an astrophysical object placed behind a given lens. This is achieved by means of the optical scalars discovered by Sachs et al. in \cite{Sachs}. Such scalars are tightly connected to the analysis of null geodesic congruences. An excellent presentation of the latter can be found in \cite{Wald} while applications to lensing are treated in \cite{Seitz}. Let us go now straight to the point and present the main results. We start by the observation that, since the spin coefficient $\rho$ is real, the congruence of null geodesics is hypersurface orthogonal. In addition, the spin coefficient $\alpha$ is real and $\alpha+\beta=\tau$ as it can be evinced by direct inspection of (\ref{aa}). It was shown in \cite{CH} that under these conditions $\bm{\ell}$ is equal to the gradient of a scalar field. Let us consider equations ($310$a) and ($310$b) (see $\S$8(d) p. 46 in \cite{CH}) describing how the spin coefficients $\rho$ and $\sigma$ vary along the geodesics, namely
\begin{eqnarray}
D\rho&=&\rho^2+|\sigma|^2+\Phi_{00},\label{eins}\\
D\sigma&=&2\sigma\rho+\Psi_0,\label{zwei}
\end{eqnarray}
We recall that in the Newmann-Penrose formalism the ten independent components of the Weyl tensor are replaced by five scalar fields $\Psi_0,\cdots,\Psi_4$ while the ten components of the Ricci tensor are expressed in terms of the scalar fields $\Phi_{ab}$ with $a,b=0,1,2$ and the Ricci scalar $R$ is written by means of the scalar field $\Lambda=R/24$. Moreover, we checked with the help of equations ($310$a-$310$r) in $\S$8(d) p. 46-7 in \cite{CH} that the only non-vanishing scalar fields $\Psi_i$, $\Phi_{ab}$, and $\Lambda$ for the C-metric expressed in terms of the tetrad (\ref{new tetrad}) are
\begin{eqnarray}
\Psi_2&=&\frac{1}{3}\left[\ell^r\partial_r\mu+n^r\partial_r\gamma-m^\vartheta\partial_\vartheta(\pi+\alpha)+\overline{m^\vartheta}\partial_\vartheta\beta+(\alpha-\beta)(\alpha-\beta+\pi)\right],\\
\Phi_{11}&=&\frac{1}{2}\left[n^r\partial_r\gamma+m^\vartheta\partial_\vartheta\alpha-\overline{m^\vartheta}\partial_\vartheta\beta+\tau^2-\mu\rho-(\alpha-\beta)^2\right],\\
\Lambda&=&\Psi_2-\Phi_{11}+m^\vartheta\partial_\vartheta\alpha-\overline{m^\vartheta}\partial_\vartheta\beta-\mu\rho-(\alpha-\beta)^2.
\end{eqnarray}
Taking into account that the optical scalars $\theta$ and $\omega$ are defined in terms of the spin coefficients as \cite{CH}
\begin{equation}
\theta=-\Re{\rho},\quad \omega=\Im{\rho},
\end{equation}
and observing that in our case $\omega=0$, equations (\ref{eins}) and (\ref{zwei}) can be rewritten as follows
\begin{eqnarray}
D\theta&=&-\theta^2-|\sigma|^2-\Phi_{00},\label{drei}\\
D\sigma&=&-2\sigma\theta+\Psi_0,\label{vier}
\end{eqnarray}
where $D=\ell^a\partial_a$. Let us check that the system of equations (\ref{drei}) and (\ref{vier}) is consistent. First of all, a vanishing spin coefficient $\sigma$ in equation (\ref{vier}) implies that $\Psi_0=0$. On the other hand, $\rho$ being real leads to $\theta=-\rho$ so that (\ref{drei}) gives
\begin{equation}\label{hoch}
\Phi_{00}=\ell^r\partial_r\rho-\rho^2=\frac{3}{8F^4}\left(\frac{\partial F}{\partial r}\right)^2-\frac{1}{4F^3}\frac{\partial^2 F}{\partial r^2}=0,
\end{equation}
where we replaced $F(r,\theta)=(r+i\alpha\cos{\vartheta})^{-2}$ in the last equality in (\ref{hoch}). Hence, we conclude that the Sachs optical scalars for the $C$-metric expressed in terms of the tetrad (\ref{new tetrad}) are
\begin{equation}\label{ergebnis}
\theta=-\frac{1}{r\sqrt{2F}}=-\frac{1}{\sqrt{2}r}-\frac{\alpha}{\sqrt{2}}\cos{\vartheta},\quad\omega=\sigma=0.
\end{equation}
At this point some comments are in order. The interpretation of the results obtained in (\ref{ergebnis}) becomes clear if we recall some basic facts related to the optical scalars. More precisely, if $N$ is a null ray, then the vector $\bm{\ell}$ which is tangent to $N$, and the real part of $\bm{m}$ which is a complex vector orthogonal to $\bm{\ell}$, will span a two-dimensional subspace (i.e. a plane) at a point $P$ on $N$. Let $\Pi$ denote such a plane. Furthermore, we take a small circle having centre at $P\in N$ and belonging to the plane $\Pi$ which is orthogonal to $\bm{\ell}$. Let us imagine to go together with the rays of the congruence $\bm{\ell}$ intersecting the circle into the future null-direction. What happens to the circle? In general, it may undergo a contraction/expansion, a rotation, and/or a shearing transformation. In particular, the optical scalar $\theta$ measures the contraction/expansion, $\omega$ quantifies the twist/rotation of the light beam, and $|\sigma|$ is the magnitude of the shear deforming the original circle into an ellipse. First of all, it is gratifying to observe that in the case of the $C$-metric there is no rotation, i.e. $\omega=0$, and this result is in line with the fact that our congruence is hypersurface orthogonal. The vanishing shear, i.e. $\sigma=0$, is consistent with the $C$-metric being algebraically special and of Petrov type-D, and hence, the Goldberg-Sachs theorem ensures a priori the existence of congruences generated by the two null-directions $\bm{\ell}$ and $\bm{n}$ such that they are both geodesics and shear-free, that is $\kappa=\sigma=\nu=\lambda=0$ and $\Psi_0=\Psi_1=\Psi_3=\Psi_4=0$. Regarding the compression factor $\theta$ found in (\ref{ergebnis}), we observe that $\theta<0$. This indicates that the radius of the light beam with small circular cross section with the plane $\Pi$ gets compressed as it propagates through the gravitational field of a $C$-black hole. Furthermore, our formula for $\theta$ predicts that on the equatorial plane of the $C$-black hole, i.e. $\vartheta=\pi/2$, the compression factor is the same as the one obtained in the Schwarzschild case. As a consequence, images of sources lying on the equatorial plane would experience the same degree of compression no matter if the lens is represented by a Schwarzschild black hole or by a black hole pulled by a cosmic string. With respect to this particular feature, we may say that lensing by a $C$-black hole is indistinguishable from lensing by a Schwarzschild black hole (see also equation (\ref{effect})). This may lead us to think that searching experimentally for $C$-black holes should focus on source images not appearing aligned with the equatorial plane of the lens. In the following, we give a simple line of reasoning explaining why it may not be impossible to distinguish a Schwarzschild black hole from a $C$-black hole by probing into effects in the optical scalar $\theta$. To this purpose, let $\theta_{H,S}$ and $\theta_{H,C}$ denote the compressions at the event horizon for a Schwarzschild and a $C$-black hole, respectively. Then,  (\ref{ergebnis}) implies
\begin{equation}\label{effect}
\left|\theta_{H,C}-\theta_{H,S}\right|=\frac{\alpha}{\sqrt{2}}|\cos{\vartheta}|<\frac{\alpha}{\sqrt{2}},
\end{equation}
that is, the effect on $\theta$ due to a $C$-black hole is at most of order $\alpha$, i.e. of the order of the acceleration parameter. On the other side, the acceleration parameter must satisfy the condition $\alpha<1/(2M)$ from which it follows that $r_h>r_H$. Physically interesting $C$-black holes will be those for which $r_h\gg r_H$. In particular, we consider the scenario where the acceleration horizon has a size comparable with $r_u$ the radius of the observable universe, that is we assume $r_h\approx r_u=4.4\cdot 10^{26}$ m \cite{gott}. Restoring SI units yields for the acceleration horizon
\[
r_h=\frac{c^2}{\alpha}.
\]
Its extension will be of the order of the horizon of the visible universe if the given $C$-black hole has an acceleration $\alpha\approx 2\cdot 10^{-10}$ m/s$^2$. Switching from geometrized units to SI units the dimensionless parameter $x=\alpha M$ reads
\begin{eqnarray}
\widetilde{x}=\frac{G}{c^4}M\alpha,
\end{eqnarray}
where $G$ denotes Newton gravitational constant.
\begin{table}[ht]
\caption{Numerical values of the upper bound for $\left|\theta_{H,C}-\theta_{H,S}\right|$ given in (\ref{effect}) for different black hole scenarios. Here, $M_\odot=1.989\cdot 10^{30}$ Kg denotes the solar mass.}
\begin{center}
\begin{tabular}{ | l | l | l | l|l|l|}
\hline
$\mbox{Lens name}$ & $M/M_\odot$  & $\alpha$ (m/s$^2$) & $\widetilde{x}=G\alpha M/c^4$  & $\left|\theta_{H,C}-\theta_{H,S}\right|$\\ \hline
\mbox{TON618}                & $6.6\cdot 10^{10}$ & $2.2\cdot 10^{-10}$  & $2.4\cdot 10^{-13}$ & $<10^{-10}$\\ \hline
\mbox{Sagittarius A$^{*}$}   & $4.3\cdot 10^{6}$  & $2.2\cdot 10^{-10}$  & $1.5\cdot 10^{-17}$ & $<10^{-10}$  \\ \hline
\mbox{GW170817}              & 2.74               & $2.2\cdot 10^{-10}$  & $9.8\cdot 10^{-24}$ & $<10^{-10}$  \\ \hline
\end{tabular}
\label{tableEins}
\end{center}
\end{table}
From equation (\ref{effect}), we observe that the effect on the image compression due to the acceleration of the black hole is maximized when $|\cos{\vartheta}|$ is close to one. This means that images of sources appearing close to the rays $\vartheta=0$ and $\vartheta=\pi$, i.e. along the cosmic string causing the black hole to accelerate, are the most promising candidates to be studied. Numerical values of the upper bound appearing in (\ref{effect}) are presented in Table~\ref{tableEins} for different black hole scenarios.

\section{WEAK AND STRONG GRAVITATIONAL LENSING IN THE C-METRIC}
We consider a gravitational field represented by the line element (\ref{line-el01}). The corresponding relativistic Kepler problem for light propagating in a C-manifold has been already discussed in Section~\ref{RKP}. A key observation allowing to study the weak lensing in the present problem by a method similar to that used in \cite{usPRD} for a Schwarzschild-de Sitter black hole, is that the critical point(s) of the effective potential (\ref{Ueff}) coincides with the critical point(s) of the corresponding geodesic equation as it was already observed in Section~\ref{RKP}. In particular, the critical point $(r_c,\vartheta_c)$ of the effective potential represented by
(\ref{crit_point_r}) and (\ref{crit_point_theta}) is not only a saddle point of the effective potential but also a critical point of the dynamical system (\ref{g7}-\ref{g8}). Due to the restriction $\alpha M\in(0,1/2)$, the angular variable associated to the critical points cannot take on any value from $0$ to $\pi$ but it is constrained to the range $(\vartheta_{c,min},\pi/2)$ with $\vartheta_{c,min}=\arccos{(1/3)}\approx 70.52^{\circ}$. At this point if we fix $\vartheta=\vartheta_c$, (\ref{g8}) is trivially satisfied and a straightforward computation shows that (\ref{g7}) can be brought into the form
\begin{equation}\label{vorher}
A(r,\vartheta_c)\left(\frac{dr}{d\lambda}\right)^2-\frac{\mathcal{E}^2}{B(r,\vartheta_c)}+\frac{\ell^2}{D(r,\vartheta_c)}=0.
\end{equation}
If we multiply the above equation by $1/2$, we can easily express (\ref{vorher}) as
\begin{equation}
\frac{A(r,\vartheta_c)B(r,\vartheta_c)}{2}\left(\frac{dr}{d\lambda}\right)^2+U_{eff}(r,\vartheta_c)=E,
\end{equation}
which coincides with (\ref{cos}) in the case $\vartheta=\vartheta_c$. At this point, we can attack the weak lensing problem as in the Schwarzschild-de Sitter case but with the cosmological horizon replaced now by the acceleration horizon. The radial equation (\ref{vorher}) can be seen as the most important equation of motion since the angular motion is completely specified by (\ref{g6}) and the condition $\vartheta=\vartheta_c$ whereas the connection between $t$ and $\lambda$ is fixed by (\ref{g5}). If we integrate (\ref{vorher}) once more, we obtain $r=r(\lambda)$, and if we substitute this function into (\ref{g5}) and (\ref{g6}), we get after integration $t=t(\lambda)$ and $\phi=\phi(\lambda)$. Elimination of the parameter $\lambda$ yields $r=r(t)$ and $\phi=\phi(t)$. Together with $\vartheta=\vartheta_c$ they represent the full solution of the problem. The involved integrals cannot be in general solved in terms of elementary functions. We want to determine the trajectory $\phi=\phi(r)$ when $\vartheta=\vartheta_c$. First of all, we observe that (\ref{vorher}) gives
\begin{equation}\label{25.19}
\left(\frac{dr}{d\lambda}\right)^2=\frac{1}{A(r,\vartheta_c)}\left[\frac{\mathcal{E}^2}{B(r,\vartheta_c)}-\frac{\ell^2}{D(r,\vartheta_c)}\right].
\end{equation}
Taking into account that $d\phi/d\lambda=(d\phi/dr)(dr/d\lambda)$ and using (\ref{g6}) together with (\ref{25.19}), we obtain
\begin{equation}\label{25.20}
\left(\frac{d\phi}{dr}\right)^2=\frac{A(r,\vartheta_c)B(r,\vartheta_c)}{D(r,\vartheta_c)}\left[\frac{\mathcal{E}^2}{\ell^2}D(r,\vartheta_c)-B(r,\vartheta_c)\right]^{-1}
\end{equation}
and integration yields
\begin{equation}\label{25.21}
\phi(r)=\pm\int~dr\sqrt{\frac{A(r,\vartheta_c)B(r,\vartheta_c)}{D(r,\vartheta_c)}}
\left[\frac{\mathcal{E}^2}{\ell^2}D(r,\vartheta_c)-B(r,\vartheta_c)\right]^{-1/2}+\widetilde{c},
\end{equation}
where $\widetilde{c}$ is an arbitrary integration constant. The plus and minus sign must be chosen for a light beam approaching the gravitational object such that its trajectory exhibits an anticlockwise and clockwise direction, respectively. This integral determines the trajectory $\phi=\phi(r)$ when $\vartheta=\vartheta_c$ and it depends essentially on one parameter, namely the ratio $\mathcal{E}/\ell$ that can be interpreted as an impact parameter $b$ as follows \cite{Weinberg}
\begin{equation}\label{impact}
\frac{1}{b^2}=\frac{\mathcal{E}^2}{\ell^2}=\frac{B(r_0,\vartheta_c)}{D(r_0,\vartheta_c)}=\frac{1}{\kappa^2 g(\vartheta_c)\sin^2{\vartheta_c}}\frac{f(r_0)}{r_0^2},
\end{equation}
where $r_0>2M$ is the distance of closest approach. This in turn permits to express (\ref{25.21}) as \cite{Virba1}
\begin{equation}
\phi(r)=\pm\int~dr\sqrt{\frac{A(r,\vartheta_c)}{D(r,\vartheta_c)}}
\left[\frac{B(r_0,\vartheta_c)D(r,\vartheta_c)}{B(r,\vartheta_c)D(r_0,\vartheta_c)}-1\right]^{-1/2}+\widetilde{c}.
\end{equation}
At this point a couple of remarks are in order. As a validity check of (\ref{impact}), let $r_0=r_c$ where $r_c$ denotes the radius of the circular orbits. Then, the corresponding value of the impact parameter also known as the critical impact parameter, here denoted by $b_c$, can be obtained from (\ref{impact}) as
\begin{equation}\label{imp}
b_c=\kappa\sin{\vartheta_c}\sqrt{g(\vartheta_c)}\frac{r_c}{\sqrt{f(r_c)}}.
\end{equation}
Let us recall that in the case of vanishing acceleration, i.e. $\alpha\to 0$, the $C$-metric goes over into the Schwarzschild metric. A Taylor  expansion of (\ref{imp}) around $\alpha=0$  leads to
\begin{equation}\label{exp}
b_c=3\sqrt{3}M-6\sqrt{3}M^2\alpha+27\sqrt{3}M^2\alpha^2+\mathcal{O}(\alpha^3).
\end{equation}
It is gratifying to observe that (\ref{exp}) correctly reproduces the Schwarzschild critical impact parameter in the limit $\alpha\to 0$.  Furthermore, depending on the values of the impact parameter we have the following scenarios
\begin{enumerate}
\item
if $b<b_c$, the photon is doomed to be absorbed by the black hole;
\item
if $b>b_c$, the photon will be deflected. Here, we must consider two further cases
\begin{enumerate}
\item
if $b\gg b_c$, the orbit is almost a straight line and we have weak gravitational lensing, i.e. the distance of closest approach is much larger than the radius $r_c$ of the photon circular orbit.
\item
If $0<b\lessapprox b_c$, we are in the regime of strong gravitational lensing corresponding to a distance of closest approach $r_0\approx r_c$. In this case the photon can orbit several times around the black hole before it flies off.
\end{enumerate}
\end{enumerate}
\begin{figure}\label{bc}
\includegraphics[scale=0.35]{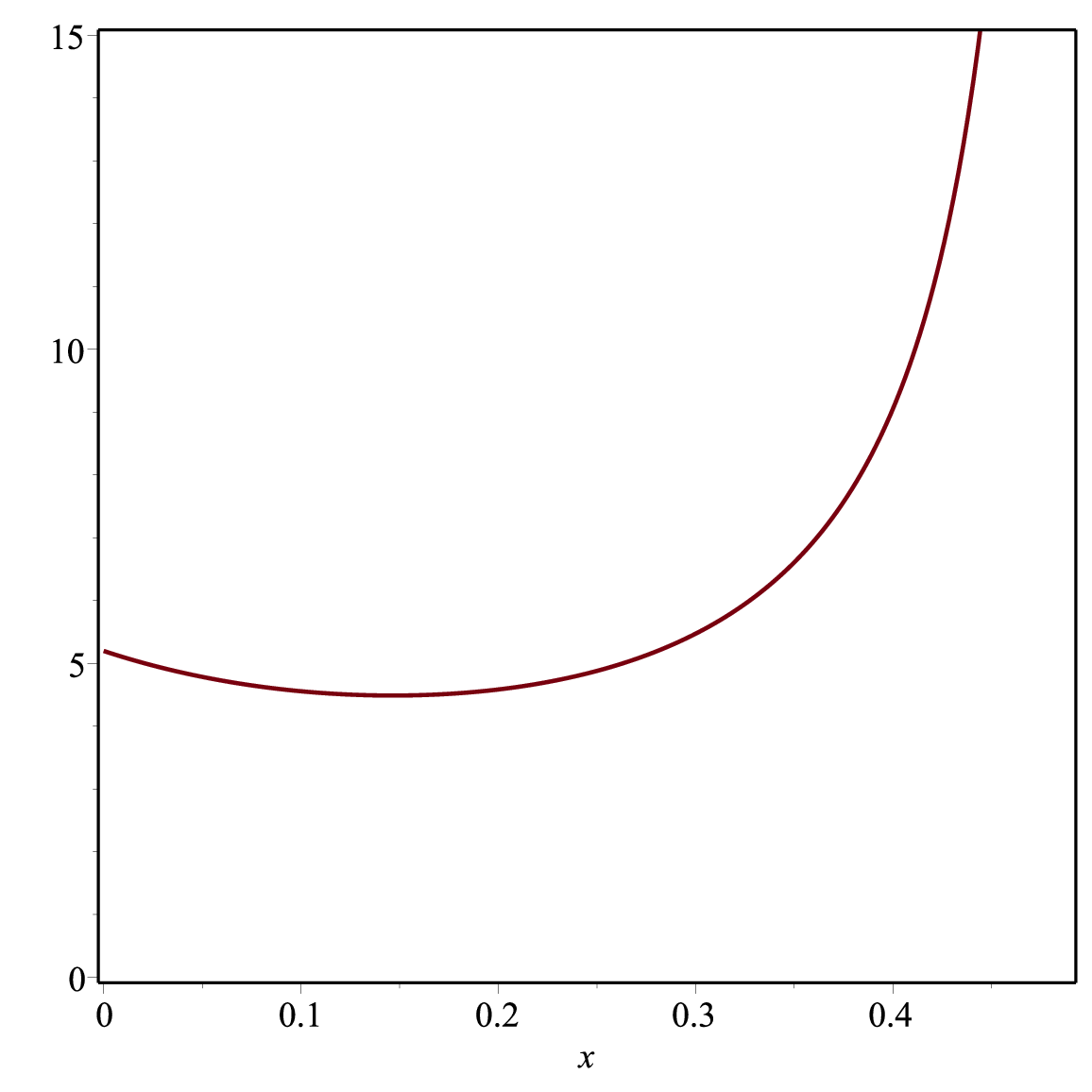}
\caption{\label{fig09}
Plot of the critical impact parameter per unit black hole mass $b_c/M$ defined in (\ref{imp}) as a functions of $x=\alpha M$ on the interval $(0,1/2)$. Note that $b_c/M$ reduces to the corresponding Schwarzschild value $3\sqrt{3}$ for alpha $x=0$ and it exhibits a minimum at $x_m=\frac{1}{2}-\frac{\sqrt{2}}{4}$ where $b_c/M=4\sqrt{3}(3+2\sqrt{2})/9$.}
\end{figure}
In order to proceed further with the analysis of the weak gravitational lensing, we rescale the radial coordinate as $\rho=r/r_H$. Then, the  function $f$ can be rewritten in terms of only one dimensionless parameter $\widehat{x}=2\alpha M$ as
\[
f(\rho)=1-\frac{1}{\rho}-{\widehat{x}}^2 (\rho^2-\rho)
\]
with $0<\widehat{x}<1$. Note that the rescaling introduced above will map the event and acceleration horizons, $r_H$ and $r_h$, to $\rho_H=1$ and $\rho_h=1/\widehat{x}>1$, respectively. The radius of the circular orbits is instead given by 
\begin{equation}\label{crit_resc}
\rho_c=\frac{r_c}{r_H}=\frac{3}{1+\sqrt{1+3\widehat{x}^2}}
\end{equation}
and in the limit of $\widehat{x}\to 0$ it correctly reproduces the radius of the photon sphere in the Schwarzschild metric, that is $\rho_\gamma=3/2$. Using formula (\ref{25.20}) yields
\begin{equation}\label{VIX}
\frac{d\phi}{d\rho}=\pm\frac{1}{\kappa\sqrt{g(\vartheta_c)}\sin{\vartheta_c}}\frac{1}{\rho^2}
\frac{1}{\sqrt{\frac{\kappa^2 g(\vartheta_c)r^2_H\sin^2{\vartheta_c}}{b^2}-\frac{1}{\rho^2}
\left[1-\frac{1}{\rho}-\widehat{x}^2(\rho^2-\rho)\right]}}.
\end{equation}
Taking into account that
\begin{equation}
\frac{\kappa^2 g(\vartheta_c)r^2_H\sin^2{\vartheta_c}}{b^2}=\frac{f(\rho_0)}{\rho_0^2}>0,
\end{equation}
equation (\ref{VIX}) can be rewritten as
\begin{eqnarray}
\frac{d\phi}{d\rho}&=&\pm\frac{1}{\kappa\sqrt{g(\vartheta_c)}\sin{\vartheta_c}}
\frac{1}{\rho^2\sqrt{\frac{f(\rho_0)}{\rho_0^2}-\frac{f(\rho)}{\rho^2}}},\label{opla1}\\
&=&\pm\frac{\rho_0}{\kappa\sqrt{g(\vartheta_c)}\sin{\vartheta_c}}\sqrt{\frac{\rho_0}{\rho}}\frac{1}{\sqrt{(\widehat{x}^2\rho_0^2+\rho_0-1)\rho^3-\rho_0^3(\widehat{x}^2\rho^2+\rho-1)}},\label{opla2}
\end{eqnarray}
where $\rho_0$ denotes the rescaled distance of closest approach. If we look at the term under the square root in (\ref{opla2}), we realize that we need to introduce a motion reality condition represented by
\begin{equation}\label{cub2}
(\widehat{x}^2\rho_0^2+\rho_0-1)\rho^3-\rho_0^3(\widehat{x}^2\rho^2+\rho-1)\geq 0.
\end{equation}
Turning points will be represented by the roots of the cubic equations associated to (\ref{cub2}). We immediately see that $\rho=\rho_0$ is a root while the other two roots are given by
\begin{equation}\label{roots}
\rho_\pm=\frac{\rho_0}{2(\widehat{x}^2\rho_0^2+\rho_0-1)}\left[1-\rho_0\pm\sqrt{(4\widehat{x}^2+1)\rho_0^2+2\rho_0-3}\right].
\end{equation}
First of all, we observe that the denominator in (\ref{roots}) can never vanish because the conditions $0<\widehat{x}<1$ and $\rho_0>1$ leads to the estimate $\rho_0-1<\widehat{x}^2\rho_0^2+\rho_0-1<\rho_0^2+\rho_0-1$ where $\rho_0^2+\rho_0-1>1$ because $\rho_0>1$. Furthermore, the roots $\rho_\pm$ are real because $(4\widehat{x}^2+1)\rho_0^2+2\rho_0-3>4\widehat{x}^2+1+2-3=4\widehat{x}^2>0$ since $\widehat{x}>0$. This shows that the cubic equation (\ref{cub2}) admits three real turning points. What is the ordering of the points $\rho_H$, $\rho_0$, $\rho_\pm$, and $\rho_h$? This question is relevant for the computation of the integral giving the deflection angle. Going back to (\ref{roots}) it is easy to see that $\rho_{-}<0$ because $1-\rho_0<0$. Clearly, $\rho_0$ is positive. Furthermore, $\rho_+>0$ because rationalizing (\ref{roots}) yields
\begin{equation}\label{rplus}
\rho_{+}=\frac{2\rho_0}{\sqrt{(4\widehat{x}^2+1)\rho_0^2+2\rho_0-3}+\rho_0-1},
\end{equation}
where $\rho_0>\rho_H=1$. The study of the turning points which is of utmost importance to analyse the motion reality condition leads to the following scenarios that we display here below. The details of the computations can be found in Appendix~\ref{TP}.
\begin{enumerate}
\item
In the case of the weak gravitational lensing, $\rho_0\gg 1$. This implies that the distance of closest approach will definitely exceed the radius $\rho_\gamma$ of the Schwarzschild photon sphere which is an upper bound for the radius of the null circular orbits in the $C$-metric. In general, for $\rho_0>\rho_\gamma>\rho_c$, it turns out that $\rho_+<\rho_c<\rho_0$ for any $\widehat{x}\in(0,1)$. Hence, the order of the roots is $\rho_{-}<0<\rho_+<\rho_0$ and the cubic in (\ref{cub2}) will be nonnegative in the interval $[\rho_0,\rho_h)$.
\item
If $\widehat{x}=x_p=\sqrt{3-2\rho_0}/\rho_0$, there are only two turning points, namely $\rho_-<0$ and $\rho_0=\rho_+=\rho_c$. In this case, the cubic (\ref{cub2}) is nonnegative on the interval $[\rho_c,\rho_h)$.
\item
In the case of strong gravitational lensing, we must impose $\rho_c\lesssim\rho_0<\rho_\gamma$. More precisely, the condition $\rho_c\lesssim\rho_0$ is satisfied if and only if $\widehat{x}\in(x_p,1)$. For such values of $x$, it is possible to show that  $\rho_+<\rho_0$. As a consequence, the cubic in (\ref{cub2}) will be nonnegative in the interval $[\rho_0,\rho_h)$. The parameter $\widehat{x}$ restricted to the same interval also ensures that $\rho_+<\rho_c$.
\end{enumerate}
 Let $\rho_{b}$ denote the position of the observer. According to the above discussion we must take $\rho_0$ and $\rho_b$ in the interval $(\rho_c,\rho_h)$. Then, (\ref{opla1}) together with the relation $\phi=\varphi/\kappa$ can be rewritten as
\begin{equation}\label{soluza}
\varphi(\rho_0)=\frac{1}{\sqrt{g(\vartheta_c)}\sin{\vartheta_c}}\int_{\rho_0}^{\rho_b}\frac{d\rho}{\rho\sqrt{f(\rho)}}\left[\left(\frac{\rho}{\rho_0}\right)^2 \frac{f(\rho_0)}{f(\rho)}-1\right]^{-1/2}.
\end{equation}
The main difference with the Schwarzschild case is that the observer cannot be positioned asymptotically in a region where we can assume that the space-time is described by the Minkowski metric. For this reason we will proceed as in \cite{usPRD} and suppose that the deflection angle is given by the formula
\begin{equation}\label{formula}
\Delta\varphi(\rho_0)=\kappa_1 I(\rho_0)+\kappa_2,
\end{equation}
where $I(\rho_0)$ denotes the r.h.s. of equation (\ref{soluza}) and $\kappa_1$, $\kappa_2$ are two scalars to be determined in such a way that the weak field approximation of (\ref{formula}) in the limit of vanishing acceleration reproduces the weak field approximation for the Schwarzschild case. Let $\widehat{\rho}=\rho/\rho_0$. Then, we have
\begin{equation}\label{integral}
I(\rho_0)=\frac{1}{\sqrt{g(\vartheta_c)}\sin{\vartheta_c}}\int_1^{\widehat{\rho}_b}\frac{d\widehat{\rho}}{\widehat{\rho}\sqrt{\widehat{\rho}^2 f(\rho_0)-f(\rho_0\widehat{\rho})}}.
\end{equation}
Instead of approximating the function $f(\rho)$ we rewrite the function under the square root in (\ref{integral}) as follows
\begin{equation}\label{integrand}
\widehat{\rho}^2 f(\rho_0)-f(\rho_0\widehat{\rho})=\frac{1}{\epsilon}\left[\epsilon^2\left(\frac{1}{\widehat{\rho}}-\widehat{\rho}^2\right)+\epsilon\left(\widehat{\rho}^2-1\right)+\widehat{x}^2\left(\widehat{\rho}^2-\widehat{\rho}\right)\right],\quad\epsilon=\frac{1}{\rho_0}.
\end{equation}
It is interesting to observe that (\ref{integral}) will indeed depend on the acceleration parameter $\alpha$ through the constant term multiplying the integral and the parameter $\widehat{x}=2\alpha M$ appearing in  (\ref{integrand}). Hence, the present situation is fundamentally different from the one occurring in the case of a Schwarzschild-de Sitter black hole where the cosmological constant  $\Lambda$ has an influence over the orbits of massive particles but it can be made to disappear from the coordinate orbital equation when photons are considered \cite{Islam}. In order to construct a perturbative expansion of (\ref{integral}), let us rewrite the integrand in (\ref{integrand}) as follows
\begin{equation}\label{FF}
F(\widehat{\rho},\epsilon,\mu)=\frac{1}{\widehat{\rho}\sqrt{\widehat{\rho}^2-1+\epsilon\left(\frac{1}{\widehat{\rho}}-\widehat{\rho}^2\right)+\mu\left(\widehat{\rho}^2-\widehat{\rho}\right)}},\quad\mu=\frac{\widehat{x}^2}{\epsilon}.
\end{equation}
At this point a remark is in order. The acceleration parameter $x$ entering in the above expression is extremely small for the class of black holes considered in Table~\ref{tableEins}. Indeed, switching from $x$ expressed in geometrized units back to the corresponding expression in SI units, here denoted by $\widehat{x}$, we have $\widehat{x}\lessapprox 10^{-13}$. Regarding the parameter $\epsilon$, we know that the deflection angle experienced by a light ray just grazing the Sun is $\Delta\varphi_{Sun}\approx 8.41\cdot 10^{-6}$ rad \cite{Weinberg}. This observation suggests that we may fix the distance of closest approach $r_0$ and hence, $\rho_0$ in such a way that at the first order in $\epsilon$ we have $\Delta\varphi_{TON618,S}\approx\Delta\varphi_{Sun}$ where the subindex $S$ stands for Schwarzschild. Searching for weak lensing effects of the same order of the one observed in the case of the Sun allows to estimate the distance of closest approach according to the formula $r_0\approx R_s M/M_{\odot}$ with $R_s$ and $M_\odot$ denoting the radius and the mass of the sun, respectively. For instance, fixing the black hole mass $M$ to be of the same order as that of TON618 we find that in SI units $r_0\approx 4.6\cdot 10^{19}$ m and $\epsilon\approx 10^{-6}$ rad. This implies that $\mu\lessapprox 10^{-20}$. It is therefore legitimate to apply a perturbative method in the small parameters $\epsilon$ and $\mu$ in order to obtain an approximated formula for the deflection angle in the weak lensing regime. We can expand (\ref{FF}) according to the Taylor formula  for multivariate functions
\begin{equation}
F(\widehat{\rho},\epsilon,\mu)=\sum_{n=0}^\infty\sum_{m=0}^\infty\frac{1}{n!m!}\left.\left(\frac{\partial^{n+m} F}{\partial\epsilon^n\partial\mu^m}\right)\right|_{\epsilon=0=\mu}\epsilon^n\mu^m
\end{equation}
to find
\begin{equation}
F(\widehat{\rho},\epsilon,\mu)=f_0(\widehat{\rho})+f_1(\widehat{\rho})\epsilon+f_2(\widehat{\rho})\epsilon^2+f_3(\widehat{\rho})\epsilon^3+g_1(\widehat{\rho})\mu+f_4(\widehat{\rho})\epsilon^4+\mathcal{O}(\epsilon\mu)
\end{equation}
with
\begin{eqnarray}
f_0(\widehat{\rho})&=&\frac{1}{\widehat{\rho}\sqrt{\widehat{\rho}^2-1}},\quad
f_1(\widehat{\rho})=\frac{\widehat{\rho}^2+\widehat{\rho}+1}{2\widehat{\rho}^2(\widehat{\rho}+1)\sqrt{\widehat{\rho}^2-1}},\quad
f_2(\widehat{\rho})=\frac{3(\widehat{\rho}^2+\widehat{\rho}+1)^2}{8\widehat{\rho}^3(\widehat{\rho}+1)^2\sqrt{\widehat{\rho}^2-1}},\\
f_3(\widehat{\rho})&=&\frac{15(\widehat{\rho}^2+\widehat{\rho}+1)^3}{48\widehat{\rho}^4(\widehat{\rho}+1)^3\sqrt{\widehat{\rho}^2-1}},\quad
g_1(\widehat{\rho})=-\frac{1}{2(\widehat{\rho}+1)\sqrt{\widehat{\rho}^2-1}},\quad
f_4(\widehat{\rho})=\frac{105(\widehat{\rho}^2+\widehat{\rho}+1)^4}{384\widehat{\rho}^5(\widehat{\rho}+1)^4\sqrt{\widehat{\rho}^2-1}}.
\end{eqnarray}
Moreover, if we take into account that $x^2=\mu\epsilon$, we can expand the factor in front of the integral in (\ref{integral}) as
\begin{equation}
\frac{1}{\sqrt{g(\vartheta_c)}\sin{\vartheta_c}}=1-\frac{\widehat{x}^2}{8}+\mathcal{O}(\widehat{x}^4)=1+\mathcal{O}(\epsilon\mu).
\end{equation}
Hence, the integral (\ref{integral}) admits the expansion
\begin{equation}\label{Ialpha}
I(\rho_0)=A_0+A_1\epsilon+A_2\epsilon^2+A_3\epsilon^3+B_1\mu+A_4\epsilon^4+\mathcal{O}(\epsilon\mu),
\end{equation}
where the  coefficients in the above expression have been computed with the software Maple 18 together with the condition that $\widehat{\rho}_b\gg 1$ and they are given by the following formulae
\begin{equation}
A_0=\frac{\pi}{2},\quad
A_1=1,\quad
A_2=\frac{1}{2!}\left(\frac{15}{16}\pi-1\right),\quad
A_3=\frac{1}{3!}\left(\frac{61}{4}-\frac{45}{16}\pi\right),\quad
B_1=-\frac{1}{2},\quad A_4=\frac{1}{4!}\left(\frac{10395}{256}\pi-\frac{195}{2}\right).
\end{equation}
Hence, by means of (\ref{formula}) we find that the deflection angle is given by
\begin{equation}\label{quasi}
\Delta\varphi(\rho_0)=\kappa_1 A_0+\kappa_2+\kappa_1\left[A_1\epsilon+A_2\epsilon^2+A_3\epsilon^3+B_1\mu+A_4\epsilon^4+\mathcal{O}(\epsilon\mu)\right].
\end{equation}
In order to fix the unknown constants $\kappa_1$ and $\kappa_2$, we observe that in the limit of vanishing acceleration, (\ref{quasi}) should reproduce the deflection angle for the weak lensing in the Schwarzschild metric. Taking into account that $\mu\to 0$  as $\alpha\to 0$, comparison of the above expression with (33) in \cite{usPRD} gives $\kappa_1=2$ and $\kappa_2=-\pi$ and equation (\ref{quasi}) becomes
\begin{equation}
\Delta\varphi(\rho_0)=\frac{2}{\rho_0}+\left(\frac{15}{16}\pi-1\right)\frac{1}{\rho_0^2}+\left(\frac{61}{12}-\frac{15}{16}\pi\right)\frac{1}{\rho_0^3}-4\alpha^2 M^2\rho_0+\left(\frac{3465}{1024}\pi-\frac{65}{8}\right)\frac{1}{\rho_0^4}+\cdots.
\end{equation}
It is gratifying to observe that in the limit $\mu\to 0$ the above equation reproduces equation (33) in \cite{usPRD}. Finally, switching back to the variable $r_0$ and passing from geometric to SI units, we end up with the following formula for the deflection angle in the $C$ metric in the weak regime
\begin{equation}\label{WGL}
\Delta\varphi(r_0)=\frac{4GM}{c^2 r_0}+\left(\frac{15}{4}\pi-4\right)\frac{G^2M^2}{c^4 r_0^2}+\left(\frac{122}{3}-\frac{15}{2}\pi\right)\frac{G^3M^3}{c^6 r_0^3}-\frac{2G\alpha^2 Mr_0}{c^6}+\left(\frac{3465}{64}\pi-130\right)\frac{G^4M^4}{c^8 r_0^4}
+\cdots.
\end{equation}
From the above expression we see that the effect due to the acceleration of the black hole is represented by the term
\begin{equation}
\delta_\alpha\varphi=-\frac{2G\alpha^2 Mr_0}{c^6}.
\end{equation}
At this point, a simple computation shows that $|\delta_\alpha\varphi|\approx 10^{-20}$ rad. In the case of a less massive black holes like Sagittarius A$^{*}$, in order to produce a deflection angle of the order of $\Delta\varphi_{Sun}$, we need to take a distance of closest approach $r_0$ comparable to $r_u$ and the corresponding value of $\delta_\alpha\varphi$ is still extremely small. Hence, we conclude that a weak lensing analysis applied to a supermassive black hole like TON618 or to even lighter black holes cannot discriminate whether the astrophysical object is represented by a C-metric or a Schwarzschild metric.

To study the strong gravitational lensing in a C-manifold, we first show that the integral (\ref{integral}) can be solved exactly in terms of an incomplete elliptic function of the first kind. This will allows us to apply an asymptotic formula derived by \cite{carlson} in the case when the sine of the modular angle and the elliptic modulus both tends to one. The same technique has been already successfully used in \cite{usPRD} to obtain the Schwarzschild deflection angle in the strong field limit with a higher order of precision than the corresponding formulae derived in \cite{Bozza,Darwin,Bozza1}. By means of (\ref{soluza}) and (\ref{formula}) with $\kappa_1$ and $\kappa_2$ fixed by the weak lensing analysis we can write the deflection angle as
\begin{equation}\label{Kontovel}
\Delta\varphi(\rho_0)=-\pi+\frac{2}{\sqrt{g(\vartheta_c)}\sin{\vartheta_c}}\int_{\rho_0}^{\rho_b}\frac{d\rho}{\rho^2}\frac{1}{\sqrt{\frac{f(\rho_0)}{\rho_0^2}-\frac{f(\rho)}{\rho^2}}}.
\end{equation}
As a preliminary check before starting to solve the integral above, we verified that in the limit of vanishing acceleration and $\rho_b\to\infty$, the above formula reproduces correctly its Schwarzschild counterpart represented by equation (30) in \cite{usPRD}. Moreover, if we compute explicitly the function appearing under the square root in (\ref{Kontovel}), i.e.
\begin{equation}\label{keiser}
\frac{f(\rho_0)}{\rho_0^2}-\frac{f(\rho)}{\rho^2}=\frac{1}{\rho^3}\left[\frac{1}{\rho_0^2}\left(\rho_0\widehat{x}^2+1-\frac{1}{\rho_0}\right)(\rho-\rho_0)^3+\frac{3}{\rho_0}\left(\frac{2}{3}\rho_0\widehat{x}^2+1-\frac{1}{\rho_0}\right)(\rho-\rho_0)^2+\left(\rho_0\widehat{x}^2+2-\frac{3}{\rho_0}\right)(\rho-\rho_0)\right]
\end{equation}
and we replace the acceleration parameter $\widehat{x}$ with the rescaled critical radius $\rho_c$ according to the formula
\begin{equation}\label{xpar}
\widehat{x}^2=\frac{2}{\rho_c^2}\left(\frac{3}{2}-\rho_c\right)
\end{equation}
obtained from (\ref{crit_resc}), then the coefficient going with the term $\rho-\rho_0$ in (\ref{keiser}) takes the form
\begin{equation}
\rho_0\widehat{x}^2+2-\frac{3}{\rho_0}=\frac{\rho_0-\rho_c}{\rho_c^2\rho_0}\left[(3-2\rho_c)(\rho_0-\rho_c)-2\rho_c(\rho_c-3)\right]
\end{equation}
and it becomes evident that there will be a logarithmic divergence when $\rho_0\to\rho_c$. Note that in equation (\ref{xpar}) $\widehat{x}^2$ is indeed positive because in the case of the C-metric  the critical radius $\rho_c$ is always smaller than the radius of the Schwarzschild photon sphere. At this point, eliminating the acceleration parameter in favour of $\rho_c$ allows to cast  (\ref{Kontovel}) in the form
\begin{equation}\label{zeug}
\Delta\varphi(\rho_0)=-\pi+\frac{2\rho_c\rho_0\sqrt{\rho_0}}{\sqrt{g(\vartheta_c)}\sin{\vartheta_c}}\int_{\rho_0}^{\rho_b}\frac{d\rho}{\sqrt{\rho\mathfrak{q}(\rho)}}
\end{equation}
with $\mathfrak{q}$ denoting the cubic polynomial
\begin{equation}
\mathfrak{q}(\rho)=\left[\rho_0^2(3-2\rho_c)+\rho_c^2(\rho_0-1)\right]\rho^3+\rho_0^3(2\rho_c-3)\rho^2-\rho_0^3\rho_c^2\rho+\rho_0^3\rho_c^2.
\end{equation}
The roots of the polynomial equation $\mathfrak{q}(\rho)=0$ are computed to be
\begin{equation}
\rho_1=\rho_0,\quad
\rho_{2,3}=\frac{\rho_c\rho_0\left[\rho_c(\rho_0-1)\pm\sqrt{\rho_c^2\rho_0^2-8\rho_c\rho_0^2+2\rho_c^2\rho_0+12\rho_0^2-3\rho_c^2}\right]}{2(2\rho_c\rho_0^2-\rho_c^2\rho_0-3\rho_0^2+\rho_c^2)}
\end{equation}
and if we replace (\ref{xpar}) into (\ref{roots}), we discover that $\rho_2=\rho_{-}$ and $\rho_3=\rho_+$. This implies that the analysis performed after equation (\ref{rplus}) can be carried over to the present situation and hence, in the case of strong lensing, we have the ordering $\rho_2<0<\rho_3<\rho_c\lesssim\rho_0<\rho_\gamma<\rho_b$ with $\rho_\gamma=3/2$. This allows us to factorize the polynomial $\mathfrak{q}$ as
\begin{equation}
\mathfrak{q}(\rho)=\left[\rho_0^2(3-2\rho_c)+\rho_c^2(\rho_0-1)\right](\rho-\rho_0)(\rho-\rho_2)(\rho-\rho_3).
\end{equation}
At this point, the integral in (\ref{zeug}) can be computed as in \cite{Gradsh} or equivalently, by means of Maple 18. In both cases, we end up with the following result expressing the deflection angle in terms of an incomplete elliptic integral of the first kind
\begin{equation}\label{DefAng}
\Delta\varphi(\rho_0)=-\pi+\frac{4\rho_c\rho_0 F(\phi_1,k)}{\sqrt{g(\vartheta_c)}\sin{\vartheta_c}\sqrt{(\rho_3-\rho_2)\left[\rho_0^2(3-2\rho_c)+\rho_c^2(\rho_0-1)\right]}}
\end{equation}
with amplitude $\phi_1$ and modulus $k$ given by
\begin{equation}
\sin{\phi_1}=\sqrt{\frac{(\rho_3-\rho_2)(\rho_b-\rho_0)}{(\rho_0-\rho_2)(\rho_b-\rho_3)}},\quad
k=\sqrt{\frac{\rho_3(\rho_0-\rho_2)}{\rho_0(\rho_3-\rho_2)}}.
\end{equation}
Before we further manipulate the above result, we check that (\ref{DefAng}) correctly reproduces formula (32) in \cite{usPRD} which gives the corresponding deflection angle in the Schwarzschild case. First of all, we observe that $\rho_c\to 3/2$ and $\vartheta_c\to\pi/2$ in the limit of vanishing acceleration parameter. This implies that $\sqrt{g(\vartheta_c)}\sin{\vartheta_c}\to 1$. Moreover, we also need to take the limit $\rho_b\to\infty$. This will be achieved by letting $\rho_b$ approaching the acceleration horizon $\rho_h=1/\widehat{x}$ followed by $\widehat{x}\to 0$. First of all, we find that
\begin{equation}\label{grenz1}
\lim_{\widehat{x}\to 0}(\rho_3-\rho_2)=\frac{\rho_0\sqrt{\rho_0^2+2\rho_0-3}}{\rho_0-1}
\end{equation}
and hence, we have
\begin{equation}
\lim_{\widehat{x}\to 0}\frac{4\rho_c\rho_0}{\sqrt{g(\vartheta_c)}\sin{\vartheta_c}\sqrt{(\rho_3-\rho_2)\left[\rho_0^2(3-2\rho_c)+\rho_c^2(\rho_0-1)\right]}}=\frac{4\sqrt{\rho_0}}{\sqrt[4]{\rho_0^2+2\rho_0-3}}
\end{equation}
It is gratifying to observe that the above result coincides with the coefficient $A(\rho_0)$ expressed by (32) in \cite{usPRD}. Regarding the amplitude $\phi_1$ it is possible to construct the following asymptotic expansion with respect to $\rho_b$, namely
\begin{equation}
\sin{\phi_1}=\sqrt{\frac{\rho_3-\rho_2}{\rho_0-\rho_2}}\left[1+\frac{\rho_3-\rho_0}{2\rho_b}+\mathcal{O}\left(\frac{1}{\rho_b^2}\right)\right].
\end{equation}
By means of the following limit
\begin{equation}
\lim_{\widehat{x}\to 0}(\rho_0-\rho_2)=\frac{\rho_0(3\rho_0-3+\sqrt{\rho_0^2+2\rho_0-3})}{2(\rho_0-1)}
\end{equation}
and (\ref{grenz1}) it is straightforward to verify that
\begin{equation}
\lim_{\substack{\rho_b\to\rho_h\\\widehat{x}\to 0}}\sin{\phi_1}=\sqrt{\frac{2\sqrt{\rho_0^2+2\rho_0-3}}{3\rho_0-3+\sqrt{\rho_0^2+2\rho_0-3}}}=1-\frac{2}{9}\left(\rho_0-\frac{3}{2}\right)+\frac{2}{9}\left(\rho_0-\frac{3}{2}\right)^2+\mathcal{O}\left(\rho_0-\frac{3}{2}\right)^3
\end{equation}
and it agrees with the first equation in (35) given by \cite{usPRD}. Finally, with the help of the limit
\begin{equation}
\lim_{\widehat{x}\to 0}\rho_3=\frac{\rho_0(1-\rho_0+\sqrt{\rho_0^2+2\rho_0-3})}{2(\rho_0-1)}
\end{equation}
it can be verified that
\begin{equation}
\lim_{\widehat{x}\to 0}k=\sqrt{\frac{3-\rho_0+\sqrt{\rho_0^2+2\rho_0-3}}{2\sqrt{\rho_0^2+2\rho_0-3}}}=1-\frac{4}{9}\left(\rho_0-\frac{3}{2}\right)+\frac{40}{81}\left(\rho_0-\frac{3}{2}\right)^2+\mathcal{O}\left(\rho_0-\frac{3}{2}\right)^3
\end{equation}
and hence, it coincides with the second equation in (35) given by \cite{usPRD}. In order to apply a certain asymptotic formula for the incomplete elliptic integral of the first kind \cite{carlson}, we need to verify that both $\sin{\phi_1}$ and $k$ approach one simultaneously as $\widehat{x}\to 0$. First of all, we observe that the elliptic modulus $k$ can be expanded around the radius $\rho_c$ of the photon circular orbits as follows 
\begin{equation}\label{K}
k=1-\frac{1}{\rho_c(3-\rho_c)}(\rho_0-\rho_c)+\mathcal{O}\left((\rho_0-\rho_c)^2\right).
\end{equation}
Moreover, the modular angle $\phi_1$ admits the Taylor expansion
\begin{equation}\label{kac}
\sin{\phi_1}=1+\frac{\rho_c(1-\rho_b)+2\rho_b}{\rho_c(\rho_b-\rho_c)(\rho_c-3)}(\rho_0-\rho_c)+\mathcal{O}\left((\rho_0-\rho_c)^2\right).
\end{equation}
Since $\sin{\phi_1},~k\to 1$ as $\widehat{x}\to 0$, the following asymptotic formula for the incomplete elliptic integral of the first kind holds \cite{carlson} 
\begin{eqnarray}
(1-\theta)F(\phi_1,k)&=&\frac{\sin{\phi_1}}{4}\left[A\ln{\frac{4}{\cos{\phi_1}+\Delta}}+B\right],
\label{Carl}\\
A&=&6-(1+k^2)\sin^2{\phi_1},\quad
B=-2+(1+k^2)\sin^2{\phi_1}+\Delta\cos{\phi_1},\quad
\Delta=\sqrt{1-k^2\sin^2{\phi_1}}
\end{eqnarray}
with relative error bound $0<\theta<(3/8)\max^2\{\cos^2{\phi_1},\Delta^2\}$. This approximation improves the usual asymptotic expansion when $(k,\phi_1)\to (1,\pi/2)$, given by
\begin{equation}
F(\phi_1,k)\approx\ln{\frac{4}{\cos{\phi_1}+\Delta}}
\end{equation}
obtained by \cite{KNC}. The asymptotic formula (\ref{Carl}) developed by \cite{carlson} represents the sum of the first two terms of a uniformly convergent series whose $N$-th partial sum approximates $F(\phi_1,k)$. Such a series is obtained via the Mellin transform method and the symmetric elliptic integral
\begin{equation}
R_F(x,y,z)=\frac{1}{2}\int_0^\infty\frac{d\tau}{\sqrt{(\tau+x)(\tau+y)(\tau+z)}}
\end{equation}
which is related to the Legendre integral
\begin{equation}
F(\phi_1,k)=\int_0^{\phi_1}\frac{d\vartheta}{\sqrt{1-k^2\sin^2{\vartheta}}}
\end{equation}
through the relation $F(\phi_1,k)=\sin{\phi_1}R_F(\cos^2{\phi_1},\Delta^2,1)$. Finally, (\ref{Carl}) is obtained from $(1.16)$ in \cite{carlson} for the case $N=2$. Taking into account that  $\sqrt{g(\vartheta_c)}\sin{\vartheta_c}$ can be expressed in terms of the rescaled radius of the critical orbits as
\begin{equation}
\sqrt{g(\vartheta_c)}\sin{\vartheta_c}=\frac{\sqrt{2}(\rho_c+3)}{3\sqrt{3\rho_c}},
\end{equation}
applying (\ref{Carl}) to (\ref{DefAng}) and expanding the non logarithmic part around $\rho_0=\rho_c$ yields the following formula for the deflection angle in the strong gravitational lensing regime
\begin{equation}\label{SGL}
\Delta\varphi(\rho_0)=-\pi+h_1(\rho_c)+h_2(\rho_b,\rho_c)+h_3(\rho_c)\ln{\left(\frac{\rho_0}{\rho_c}-1\right)}+h_4(\rho_c)(\rho_0-\rho_c)+\mathcal{O}(\rho_0-\rho_c)^2
\end{equation}
with
\begin{eqnarray}
h_1(\rho_c)&=&-h_3(\rho_c)\ln{\left[8(3-\rho_c)^2\right]},\quad
h_2(\rho_b,\rho_c)=-h_3(\rho_c)\ln{\frac{\rho_b-\rho_c}{(3-\rho_c)\left[\sqrt{\rho_b(3-\rho_c)}+\sqrt{\rho_c+2\rho_b-\rho_b\rho_c}\right]^2}},\\
h_3(\rho_c)&=&-\frac{3\sqrt{6}\rho_c}{(3+\rho_c)\sqrt{3-\rho_c})},\quad
h_4(\rho_c)=-\frac{3\sqrt{6}(2\rho_c-7)}{2(\rho_c+3)(3-\rho_c)^{3/2}}
\end{eqnarray}
As a validity check of (\ref{SGL}), we show that it correctly reproduces the corresponding strong lensing formula in the Schwarzschild case as given in \cite{usPRD,Bozza1,Bozza} when the acceleration parameter vanishes. Recall that $\rho_c\to 3/2$ in such a limit . First of all we observe that
\begin{equation}
\xi_1=\lim_{\rho_c\to 3/2}h_1(\rho_c)=\ln{324}.
\end{equation}
To construct the corresponding limit of the coefficient function $h_2$, we let the position $\rho_b$ of the observer to approach the acceleration horizon $\rho_h=1/\widehat{x}$ and then, we construct the limit $\widehat{x}\to 0$ so that the observer is placed at space-like infinity as in the Schwarzschild case. This limiting process will also send $\rho_c$ to the rescaled radius $\rho_\gamma=3/2$ of the photon sphere. More precisely, we find that
\begin{equation}
\xi_2=\lim_{\rho_c\to\frac{3}{2}}\lim_{\substack{\rho_b\to\rho_h\\\widehat{x}\to 0}}h_2(\rho_b,\rho_c)
=-\lim_{\rho_c\to\frac{3}{2}}h_3(\rho_c)\lim_{\widehat{x}\to 0}\ln{\frac{1-\rho_c\widehat{x}}{(3-\rho_c)\left[\sqrt{3-\rho_c}+\sqrt{2-\rho_c+\rho_c\widehat{x}}\right]^2}}=\ln{\frac{4}{9}(7-4\sqrt{3})}
\end{equation}
We also checked that if we first take $\rho_c\to 3/2$ followed by $\rho_b\to\rho_h$ and $\widehat{x}\to 0$, the above result is not affected. Hence, $\xi_1+\xi_2=\ln{144(7-4\sqrt{3})}$ as the second term appearing in equation (37) of \cite{usPRD}. To conclude, we observe that
\begin{equation}
\lim_{\rho_c\to\frac{3}{2}}h_3(\rho_c)=-2,\quad \lim_{\rho_c\to\frac{3}{2}}h_4(\rho_c)=\frac{16}{9}.
\end{equation}
It is gratifying to observe that equation (\ref{SGL}) correctly reproduce formula (37) in \cite{usPRD} for a vanishing acceleration parameter. To the best of our knowledge, equations (\ref{DefAng}) and (\ref{SGL}) representing the general formula for the deflection angle and its approximation in the strong regime, respectively, are new, and in principle, they may be used together with observational data to confirm or disprove the existence of black holes modelled in terms of the C-metric.

\section{Conclusions and outlook}
Light bending and possible bound states of light are genuine effects of general relativity. Whereas light bending has been studied and even observed in a variety of situations, bound orbits of massless particles are an interesting phenomenon and they deserve a special attention ( see e.g. \cite{usPRD,Pedro,Flavio,Pedro1,Pedro2} and references therein). In general, it is known that a local maximum in the effective potential corresponds to an unstable circular orbit of light. However, a saddle point presents a more challenging problem and needs to be examined from case to case. In the C-metric of two black hole we indeed find such a saddle point. We perform a Jacobi analysis to probe into the stability issues of the circular orbit related to the saddle point. A careful lengthy analysis reveals that the orbit is unstable. Moreover, the study of the circular orbits allows to construct the impact parameter associated to the light scattering in the C-metric. This, in turn,  permits to define the weak and strong lensing regimes on a certain family of invariant cones.   On such surfaces, we derive the corresponding analytic formulae for the deflection angle. We find that in the weak gravitational lensing regime it is impossible to distinguish between a lens represented by an extremely massive black hole such as TON618 modelled in terms of the C-metric or by the Schwarzschild metric because the order of magnitude of the discrepancy between the predicted deflection angles is of the order $10^{-25}$ rad. This effect can not be observed because it is tremendously smaller than the deflection angle (order $10^{-8}$ rad) experienced by a light ray from a distant star skimming the Sun's surface, and hence, it is not an effect that may be used to hunt C-black holes in the universe.

To conclude, we mention that \cite{Lim} showed that the radial geodesics along the poles ($\vartheta=0$ and $\vartheta=\pi$) are unstable by performing a Schwarzschild-like stability analysis for an effective potential that depends on the radial variable only. However, the full effective potential is a non trivial function of the radial variable and the polar angle. We do not see any reason why the method we used here to show the (in)stability of circular orbits in the C-metric would not work also in the analysis of time-like geodesics. This more general method would allow to extend the findings in \cite{Lim}. If we keep in mind the following general feature, i.e. the Jacobi and Lyapunov stability analyses may predict different parameter ranges for stability and instability already when we deal with a dynamical system that depends on a parameter, then, in view of the fact that the dynamical system represented by (\ref{din1}) and (\ref{din2}) depends on more than one parameter, it would be advisable to couple the Jacobi stability analysis, which probes into the ''robustness'' of a given trajectory with the Lyapunov method outlined in \cite{saugen} whose applicability would first require to find a way of transforming the second order system (\ref{din1}) and (\ref{din2}) into a first order system. Last but not least, it may be interesting to use also  Lyapunov's direct method (LDM) which is based on Lyapunov functions. This approach in general may give results slightly different to those one obtained using the aforementioned two methods. The problem with LDM is that the use of this technique requires knowledge about the so-called Lyapunov function having certain properties, and there exists no general approach for constructing such function which can be used to define basins of attraction. The findings of these studies will be presented in a forth-coming paper dealing with the C-metric plus a positive cosmological constant.

\begin{acknowledgments}
The authors would like to express their gratitude for the valuable comments and suggestions they have been provided with by the anonymous referees.
\end{acknowledgments}

\appendix
\section{Coefficient functions appearing in (\ref{g7}) and (\ref{g8})}\label{App}
The Christoffel symbols for the metric (\ref{line-el01}) and the coefficient functions in (\ref{g7}) and (\ref{g8}) have been computed using Maple 18. The non vanishing Christoffel symbols are
\begin{eqnarray}
&&\Gamma^t_{tr}=\frac{\partial_r B}{2B},\quad
\Gamma^t_{t\vartheta}=\frac{\partial_\vartheta B}{2B},\quad
\Gamma^\phi_{r\phi}=\frac{\partial_r D}{2D},\quad
\Gamma^\phi_{\vartheta\phi}=\frac{\partial_\vartheta D}{2D},\quad
\Gamma^r_{tt}=\frac{\partial_r B}{2A},\quad
\Gamma^r_{rr}=\frac{\partial_r A}{2A},\quad
\Gamma^r_{r\vartheta}=\frac{\partial_\vartheta A}{2A},\\
&&\Gamma^r_{\vartheta\vartheta}=-\frac{\partial_r C}{2A},\quad
\Gamma^r_{\phi\phi}=-\frac{\partial_r D}{2A},\quad
\Gamma^\vartheta_{tt}=\frac{\partial_\vartheta B}{2C},\quad
\Gamma^\vartheta_{rr}=-\frac{\partial_\vartheta A}{2C},\quad
\Gamma^\vartheta_{r\vartheta}=\frac{\partial_r C}{2C},\quad
\Gamma^\vartheta_{\vartheta\vartheta}=\frac{\partial_\vartheta C}{2C},\\
&&\Gamma^\vartheta_{\phi\phi}=-\frac{\partial_\vartheta D}{2C}.
\end{eqnarray}
 Moreover, we have
 \begin{eqnarray}
 -\frac{\partial_r A}{2A}&=&\frac{f^{'}}{2f}+\alpha\sqrt{F}\cos{\vartheta},\quad
 \frac{\partial_r C}{2A}=r\frac{f\sqrt{F}}{g},\quad-\frac{\partial_\vartheta A}{A}=-2\alpha r\sqrt{F}\sin{\vartheta},\\
 \frac{\partial_\vartheta C}{2C}&=&\frac{g^{'}}{2g}-\alpha r\sqrt{F}\sin{\vartheta},\quad
 \frac{\partial_\vartheta A}{2C}=\frac{\alpha g\sin{\vartheta}}{rf}\sqrt{F},\quad
 -\frac{\partial_r C}{C}=-\frac{2}{r}\sqrt{F}
 \end{eqnarray}
 and
 \begin{eqnarray}
 -\frac{\mathcal{E}^2}{2}\frac{\partial_r B}{AB^2}+\frac{\ell^2}{2}\frac{\partial_r D}{AD^2}&=&-\frac{1}{F\sqrt{F}}\left[\frac{\mathcal{E}^2}{\sqrt{F}}\left(\frac{f^{'}}{2f}-\alpha\sqrt{F}\cos{\vartheta}\right)-\frac{\ell^2}{\kappa^2 r^2 g\sin^2{\vartheta}}\right],\\
 -\frac{\mathcal{E}^2}{2}\frac{\partial_\vartheta B}{CB^2}+\frac{\ell^2}{2}\frac{\partial_\vartheta D}{CD^2}&=&-\frac{g\sin{\vartheta}}{F\sqrt{F}}\left\{\frac{\alpha\mathcal{E}^2}{rf}-\frac{\ell^2}{\kappa^2 g\sin^2{\vartheta}}\left[\frac{1}{r^4\sqrt{F}}\left(\frac{\dot{g}}{2g\sin{\vartheta}}+\frac{\cos{\vartheta}}{\sin^2{\vartheta}}\right)+\frac{\alpha}{r^3}\right]\right\},
 \end{eqnarray}
 where prime denotes differentiation with respect to $r$ and dot stands for differentiation with respect to $\vartheta$.
 
\section{Critical points for the Schwarzschild metric}\label{AppS} 
First of all, it is straightforward to verify that the effective potential (\ref{Schw_m}) in the case of vanishing mass has a maximum at $r_\gamma=3M$ representing an unstable photon circular orbit. Next, we take another approach to get the radius of the photon sphere. Namely, we compute the critical point(s) of the geodesic equations (\ref{g7}) and (\ref{g8}) adapted to the Schwarzschild case, we impose that such point(s) also satisfy the corresponding constraint equation, and we show that there is only one critical point and it must coincide with the radius of the photon sphere. We start by observing that in the Schwarzschild case equation (\ref{g8}) reduces to a trivial identity because of the spherical symmetry of the manifold. Hence, if $(r_k,\vartheta_k)$ denotes a critical point of the Schwarzschild geodesic equation where $\vartheta_k$ can take on any value on the interval $[0,\pi]$, equation (\ref{g7}) becomes
\begin{equation}\label{v1}
-\frac{\mathcal{E}^2}{2}\frac{\widetilde{f}^{'}(r_k)}{\widetilde{f}(r_k)}+\ell^2\frac{\widetilde{f}(r_k)}{r_k^3}=0,\quad \widetilde{f}(r)=1-\frac{2M}{r}.
\end{equation}
Moreover, in the Schwarzschild case the constraint equation (\ref{cos}) evaluated at $r=r_k$ gives
\begin{equation}\label{v2}
\ell^2\frac{\widetilde{f}(r_k)}{r^2_k}=\mathcal{E}^2.
\end{equation}
We replace (\ref{v2}) into (\ref{v1}) to obtain the equation
\begin{equation}
\frac{r_k-3M}{r_k(r_k-2M)}=0
\end{equation}
from which it follows that $r_k=r_\gamma$. 

\section{Analysis of the turning points}\label{TP}
From (\ref{rplus}) we have
\begin{equation}\label{B1}
0<\frac{\rho_+}{\rho_0}=\frac{2}{\sqrt{(4\widehat{x}^2+1)\rho_0^2+2\rho_0-3}+\rho_0-1}.
\end{equation}
Observe that $(4\widehat{x}^2+1)\rho_0^2+2\rho_0-3>\rho_0^2+2\rho_0-3$. Since $\rho_0>\rho_\gamma=3/2$, it follows that $\sqrt{(4\widehat{x}^2+1)\rho_0^2+2\rho_0-3}>3/2$ and hence, $\sqrt{(4\widehat{x}^2+1)\rho_0^2+2\rho_0-3}+\rho_0-1>2$ by using again $\rho_0>3/2$. The latter inequality can be used in (\ref{B1}) to conclude that $\rho_+/\rho_0<1$. In order to show that $\rho_+<\rho_c$, we observe that
\begin{equation}\label{hilfe}
\rho_+-\rho_c=\frac{2\rho_0\sqrt{1+3\widehat{x}^2}-3\sqrt{(1+4\widehat{x}^2)\rho_0^2+2\rho_0-3}+\rho_0-3}{(1+\sqrt{1+3\widehat{x}^2})[\sqrt{(1+4\widehat{x}^2)\rho_0^2+2\rho_0-3}+\rho_0-1]}.
\end{equation}
The above expression does not vanish for any $\widehat{x}\in(0,1)$ and for any $\rho_0>\rho_\gamma$ because the zeroes 
\begin{equation}
\widehat{x}_{1,2}=\pm i\frac{\sqrt{2\rho_0-3}}{\rho_0},\quad x_{3,4}=\pm i\frac{\sqrt{\rho_0^2+2\rho_0-3}}{\rho_0}
\end{equation}
of the numerator in (\ref{hilfe}) are purely imaginary. This means that the graph of the function $\rho_+-\rho_c$ is always either above or below the square $\mathcal{S}=(0,1)\times(\rho_\gamma,\rho_h)$. Hence, if we take for instance $\widehat{x}=0.1$ and $\rho_0=1.6$, then $(\rho_+-\rho_c)(0.1,1.6)=-0.09<0$ and we conclude that for all $(\widehat{x},\rho_0)\in\mathcal{S}$ it must be $\rho_+<\rho_c$.\\
Let $\rho_0<\rho_\gamma$. The verification of $\rho_+=\rho_c=\rho_0$ whenever $\widehat{x}=x_p=\sqrt{3-2\rho_0}/\rho_0$ is trivial. We only need to set $\rho_+=\rho_c$ and to solve with respect to $\widehat{x}$.\\
To show that $\rho_+<\rho_0$ when $\rho_c<\rho_0<\rho_\gamma$,  we observe that by means of (\ref{B1}) we have
\begin{equation}
\frac{\rho_+}{\rho_0}-1=\frac{3-\rho_0-\sqrt{(4\widehat{x}^2+1)\rho_0^2+2\rho_0-3}}{\sqrt{(4\widehat{x}^2+1)\rho_0^2+2\rho_0-3}+\rho_0-1}.
\end{equation}
The left hand side of the above expression will be negative whenever $3-\rho_0<\sqrt{(4\widehat{x}^2+1)\rho_0^2+2\rho_0-3}$. In the case of strong gravitational lensing, we have $\rho_c\lesssim\rho_0$ and since $\rho_c<\rho_\gamma$, we can safely assume that $\rho_0<\rho_\gamma=3/2$. This implies that $3-\rho_0$ is always positive and from $(3-\rho_0)^2<(4\widehat{x}^2+1)\rho_0^2+2\rho_0-3$ after simplification we get $3-2\rho_0<\widehat{x}^2\rho_0^2$ from which we conclude that $\sqrt{3-2\rho_0}/\rho_0<\widehat{x}<1$. Let $x_p=\sqrt{3-2\rho_0}/\rho_0$ and introduce a small perturbation $\widehat{x}_\epsilon=x_p+\epsilon$ with $\epsilon>0$. Expanding $\rho_+$ and $\rho_c$ in power of $\epsilon$ gives 
\begin{eqnarray}
\rho_+(\widehat{x}_\epsilon)&=&\rho_0-2c\epsilon+\mathcal{O}(\epsilon^2),\quad c=\frac{\rho_0^2\sqrt{3-2\rho_0}}{3-\rho_0}>0\\
\rho_c(\widehat{x}_\epsilon)&=&\rho_0-c\epsilon+\mathcal{O}(\epsilon^2),
\end{eqnarray}
and we conclude that as soon as $\widehat{x}\in(x_p,1)$, we have $\rho_+<\rho_c<\rho_0$.

\end{document}